\documentclass[runningheads]{llncs}
\usepackage[a4paper,left=3cm,right=3cm]{geometry}

\usepackage[T1]{fontenc}
\usepackage{graphicx}
\usepackage{amsfonts}
\usepackage{amsmath}
\usepackage{algorithm}
\usepackage{algpseudocode}
\usepackage{censor,caption}
\usepackage{multirow}
 
\usepackage[T1]{fontenc}
\usepackage{graphicx,verbatim}
\usepackage{color}

\usepackage{booktabs}
\usepackage{array}
\usepackage{longtable}
\usepackage{tabularx}
\usepackage{threeparttable}
\usepackage{caption}
\usepackage{xurl}
\usepackage{amssymb}

\usepackage[colorlinks=true,linkcolor=blue,citecolor=blue,urlcolor=blue]{hyperref}

\begin{document}
\title{Perceptual implications of automatic anonymization in pathological speech}

\author{
Soroosh Tayebi Arasteh\inst{1,2,3,4,5,}$^{\ast}$ \and
Saba Afza\inst{1} \and
Tri-Thien Nguyen\inst{1,6} \and
Lukas Buess\inst{1} \and
Maryam Parvin\inst{1} \and
Tomas Arias-Vergara\inst{1} \and
Paula Andrea Perez-Toro\inst{1} \and
Hiu Ching Hung\inst{5} \and
Mahshad Lotfinia\inst{4,5} \and
Thomas Gorges\inst{1} \and
Elmar Noeth\inst{1} \and
Maria Schuster\inst{8} \and
Seung Hee Yang\inst{9} \and
Andreas Maier\inst{1}
}

\institute{
Pattern Recognition Lab, Friedrich-Alexander-Universit\"at Erlangen-N\"urnberg, Erlangen, Germany. \and
Department of Urology, Stanford University, Stanford, CA, USA. \and
Department of Radiology, Stanford University, Stanford, CA, USA. \and
Lab for AI in Medicine, RWTH Aachen University, Aachen, Germany. \and
Department of Diagnostic and Interventional Radiology, University Hospital RWTH Aachen, Aachen, Germany. \and
Institute of Radiology, University Hospital Erlangen, Erlangen, Germany. \and
Department of Foreign Language Education, Friedrich-Alexander-Universität Erlangen-Nürnberg, Erlangen, Germany. \and
Department of Otorhinolaryngology, Head and Neck Surgery, Ludwig-Maximilians-Universität München, Munich, Germany. \and
Speech \& Language Processing Lab, Friedrich-Alexander-Universität Erlangen-Nürnberg, Erlangen, Germany.
}

\maketitle 
{\footnotesize
$^{\ast}$Correspondence to: Soroosh Tayebi Arasteh (\email{soroosh.arasteh@rwth-aachen.de})
}

\begin{abstract}
Automatic anonymization is increasingly used to enable ethical sharing of clinical speech, yet its perceptual and clinical consequences remain undercharacterized. We present a human-centered evaluation of automatically anonymized pathological speech, using a structured protocol with ten native and non-native German listeners spanning clinical and signal-processing expertise. The cohort comprised 180 German speakers from Cleft Lip and Palate, Dysarthria, Dysglossia, Dysphonia, and adult and child controls (30 per group). Each original recording and its automatically-anonymized counterpart was evaluated on four tasks: zero-shot Turing-style discrimination, few-shot discrimination after brief familiarization, 5-point quality rating, and 4-point blinded clinical severity rating by a senior phoniatrician. Listeners detected anonymization at 91 $\pm$ 7\% zero-shot and 93 $\pm$ 6\% few-shot accuracy, with significant variation across disorders ($p = 0.008$) that attenuated with familiarization. Perceived quality dropped by 30 percentage points on a 0--100 scale ($p < 0.001$), reorganizing the perceived-quality hierarchy across groups. Native language modulated detectability but not quality degradation, while domain expertise modulated quality degradation but not detectability, a double dissociation between the two listener attributes; speaker sex and age produced no detectable bias. Clinical severity ratings were preserved at near-perfect agreement in Dysarthria, Dysglossia, and Dysphonia (quadratic-weighted Cohen's $\kappa$ $0.87$--$0.94$), with no recording shifting by more than one grade. Crucially, perceptual outcomes were decoupled from the standard computational privacy metric: the pathology with the strongest computational anonymization was the least perceptually conspicuous, and vice versa. These findings argue for disorder-stratified, listener-stratified, clinician-validated evaluation as the minimum standard for licensing anonymized speech for clinical use.

\end{abstract}


\section*{Introduction}

Speech is among the most diagnostically informative biosignals routinely collected in clinical practice, carrying patterned alterations of articulation, phonation, resonance, rhythm, and prosody that underpin assessment, rehabilitation, and outcome monitoring across speech and voice disorders \cite{kent1996hearing,riedhammer2023medical,pappagari20_interspeech,bayerl2022can}. Speech recordings also support automated systems for disorder detection and monitoring, but they encode biometric identity through vocal-tract resonance, pitch, spectral shape, speaking style, and linguistic habits. Sharing clinical recordings beyond the institution that collected them is therefore constrained by data-protection law and by the ethical duties owed to patients whose voices are inherently identifiable \cite{strimbu2010biomarkers,califf2018biomarker,speechbiomar,kroger2019privacy,tayebi2026privacy,tayebiarasteh23_interspeech,mohammadi2026differential}. Speaker anonymization, the suppression of speaker-identifying features while preserving the linguistic, paralinguistic, and clinically relevant content needed for downstream use, has consequently become a prerequisite for large-scale clinical speech sharing \cite{tomashenko2022voiceprivacy,patino2021speaker,srivastava2022privacy}.

Most speech-anonymization research is evaluated through automatic computational summaries \cite{fang2019speaker,khamsehashari22_spsc,Snyderxvector,NAUTSCH2019441,arasteh2024addressing,tomashenko2022voiceprivacy}. The VoicePrivacy challenges \cite{tomashenko2022voiceprivacy,tomashenko2024voiceprivacy,tomashenko2022voiceprivacy2022} and related work quantify privacy mainly through the equal error rate (EER) of automatic speaker verification (ASV) systems, while utility is usually measured through word- or phone-recognition performance, speech-recognition error, or downstream task performance \cite{tayebi2026privacy}. These benchmarks have been essential for comparing signal-processing approaches such as the McAdams coefficient method \cite{patino2021speaker,mcadams1984spectral} with neural approaches based on speaker embeddings, pseudo-speakers, or waveform resynthesis \cite{fang2019speaker,srivastava2022privacy}. Yet listener-based evaluations have largely focused on healthy or non-clinical speech \cite{fang2019speaker,khamsehashari22_spsc,Snyderxvector,srivastava2022privacy,srivastava20_interspeech,siegert2024user}, leaving unresolved whether anonymized pathological speech sounds natural, whether the transformation is perceptually obvious, and whether the apparent clinical impairment is preserved \cite{Kluinpercept,sachin2008clinical}.

This gap is important because the acoustic dimensions targeted by anonymization can overlap with diagnostic cues. In a prior large-scale study of more than 2700 native German speakers spanning Cleft Lip and Palate (CLP) \cite{millard2001different,harding1996characteristics,maier2006fully}, Dysarthria \cite{hirose1986pathophysiology}, Dysglossia \cite{schroter2005rehabilitation}, and Dysphonia \cite{sama2001clinical}, we evaluated automatic anonymization for pathological speech and showed that the McAdams coefficient method \cite{patino2021speaker,mcadams1984spectral} delivered substantial EER gains while largely preserving the diagnostic utility of downstream pathology classifiers \cite{arasteh2024addressing}. That work established a computational privacy-utility baseline, but it did not test whether humans can detect the transformation, perceive quality degradation, or hear the same severity of impairment after anonymization. These questions cannot be inferred from EER or classifier performance alone, because clinical speech assessment is inherently perceptual.

\begin{figure}[h]
\centering
\includegraphics[width=\textwidth]{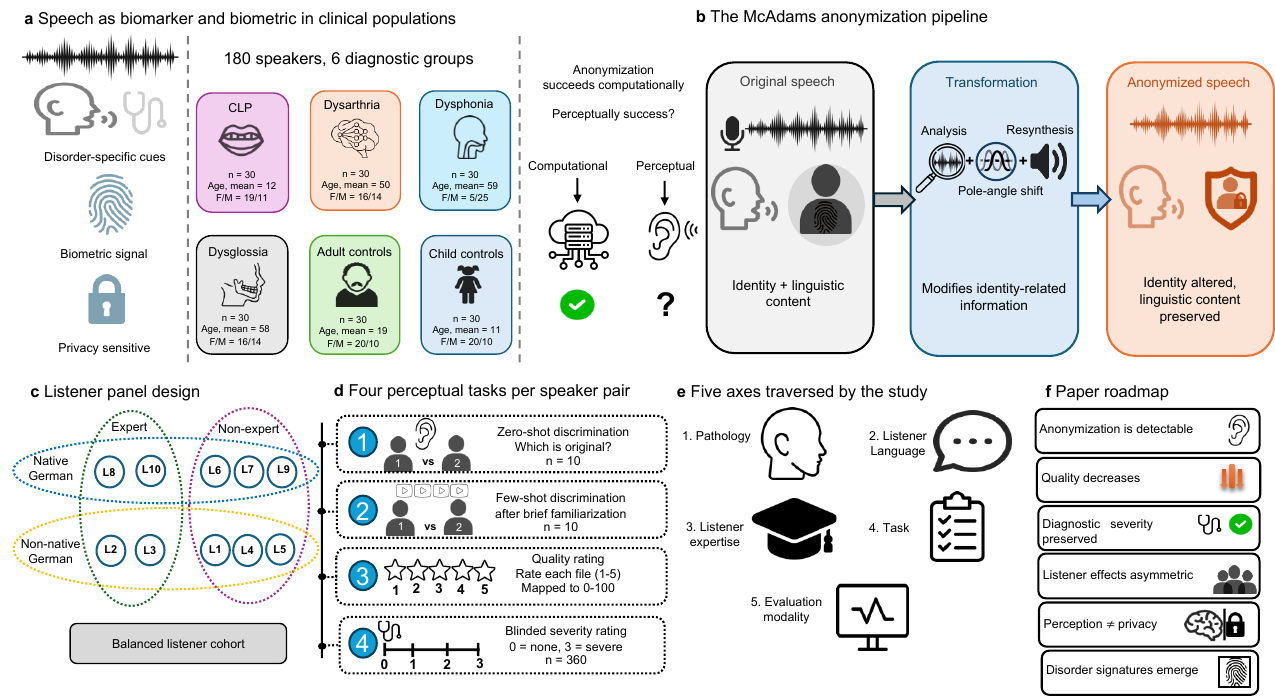}
\caption{Study overview: perceptual evaluation of speech anonymization in pathological speech.
\textbf{a} Clinical motivation and cohort: speech is simultaneously a diagnostic biomarker and a biometric identifier, creating a tension between clinical utility and privacy. The cohort comprises 180 speakers across four pathology groups (Cleft Lip and Palate (CLP), Dysarthria, Dysglossia, Dysphonia) and two control groups (adult and child), 30 per group, spanning pediatric and adult populations.
\textbf{b} The McAdams anonymization pipeline: linear-predictive analysis of the original recording, pole-angle shift toward the unit-circle origin, and resynthesis; linguistic content and source excitation are retained while speaker identity is altered.
\textbf{c} Listener panel design: 10 listeners crossing two attributes, native language (5 native German, 5 non-native) and domain expertise (4 experts with $\geq 8$ years of experience, 6 non-experts), forming a balanced factorial structure.
\textbf{d} Four perceptual tasks administered per speaker pair: zero-shot discrimination, few-shot discrimination after brief familiarization, per-file quality rating on a 5-point scale mapped to 0--100, and blinded severity rating on a 4-point ordinal scale (0 none, 3 severe) performed by a single senior phoniatrician across all 360 files.
\textbf{e} Five experimental axes traversed by the study: speaker pathology, listener native language, listener expertise, perceptual task, and evaluation modality.
\textbf{f} Roadmap of the six findings developed in the paper.}
\label{fig:overview}
\end{figure}

Here, we provide a listener-centered evaluation of automatic anonymization in pathological speech, using the McAdams coefficient method as a fixed and interpretable automatic anonymization procedure selected from our prior computational framework \cite{arasteh2024addressing} (Fig. \ref{fig:overview}). The goal is not to claim that this single method represents all anonymization systems, but to test whether computational privacy and utility evidence is sufficient once human perception is considered \cite{Kluinpercept,sachin2008clinical,Pernonpercep}. We studied 180 German-speaking patients and controls, with 30 speakers each from CLP, Dysarthria, Dysglossia, Dysphonia, adult controls, and child controls. Ten listeners completed a zero-shot Turing-style \cite{turing1950computing} discrimination task, a few-shot discrimination task after repeated exposure, and a five-point quality-rating task. To assess whether the diagnostically relevant pathology signal survives anonymization, the clinical phoniatrician in the panel additionally rated perceived pathological severity for every recording using a blinded four-point ordinal scale. The perceptual outcomes were then compared with previously reported computational privacy and utility metrics from the same clinical corpus and anonymization pipeline \cite{arasteh2024addressing}.

We expect anonymization to be detectable in clinical speech and perceived quality to decrease, in line with the spectral distortion introduced by the formant shift \cite{patino2021speaker}. We further expect these effects to vary by disorder \cite{arasteh2024addressing}, because CLP, Dysarthria, Dysglossia, and Dysphonia emphasize different acoustic substrates. The data show that listeners detect the transformation with high accuracy, that quality degradation is substantial but disorder-specific, and that perceptual outcomes are not explained by computational privacy metrics alone. Listener attributes separate across perceptual axes, whereas speaker sex and age have little effect. 
These show that clinical speech anonymization cannot be licensed by automatic privacy and utility metrics alone. Its value depends on disorder-specific perceptual consequences, the intended downstream use, and whether the clinically relevant speech abnormality remains audible after the identity-protecting transformation.

\section*{Results}

To quantify the perceptual consequences of automatic speech anonymization in clinical populations, we conducted a listener-centered study using 180 German-speaking speakers selected from the PEAKS clinical speech corpus \cite{maier2009peaks,arasteh2023effect,arasteh2024addressing}. The cohort included patients with CLP, Dysarthria, Dysglossia, and Dysphonia, as well as adult and child controls. For every speaker, one original recording and one anonymized counterpart generated with the McAdams coefficient method \cite{mcadams1984spectral,patino2021speaker,arasteh2024addressing} were retained, yielding 180 paired comparisons and 360 utterances in total. Ten listeners with mixed linguistic and professional backgrounds completed three blinded tasks: a zero-shot Turing-style discrimination task \cite{turing1950computing}, in which the original and anonymized files were presented side by side without prior exposure; a few-shot version of the same task after repeated exposure; and a five-point quality rating for each file. In addition, the senior clinical phoniatrician in the panel (M.S. with 35 years of experience) rated the perceived pathological severity of every recording on a four-point ordinal scale to assess preservation of the clinical impairment signal after anonymization. Speaker labels, diagnostic group labels, anonymization status, and within-pair file position were concealed during all listening tasks.

Table~\ref{tab:cohort} summarizes the speaker cohort and listener panel. The 180 speakers were drawn in equal numbers from four pathology groups and two control groups: CLP, Dysarthria, Dysglossia, Dysphonia, adult controls, and child controls (30 per group). The selection spans pediatric and adult populations and combines resonance-based (CLP, Dysglossia), motor-based (Dysarthria), and laryngeal (Dysphonia) etiologies. Across the full cohort, mean age was 35 $\pm$ 24 years (range 6 to 78), with 93 female and 87 male speakers. The pediatric groups (CLP and child controls) had mean ages near 11 years, while the adult pathology groups had mean ages between 50 and 59 years and varied in sex composition, most notably Dysphonia (25 male, 5 female).

The listener panel was structured by native-language background and domain expertise. Five listeners were native German speakers and five were non-native German speakers, with the latter spanning proficiencies from A1 to C1 according to the Common European Framework of Reference for Languages \cite{little2020common}. Domain expertise was defined as at least 8 years of specialized experience in either signal-level speech processing or clinical voice and speech assessment, yielding four experts (three speech-processing experts and one clinical phoniatrician) and six non-experts. Throughout the paper, listener-derived measures are reported as integer percentages, with the \% sign omitted, in the format mean $\pm$ standard deviation [95\% confidence intervals]. Statistical quantities, including p-values, correlation coefficients, and effect sizes, are reported to three decimal places.

\begin{table}[h]
\centering
\caption{Speaker cohort and listener panel. The 180 speakers span six diagnostic groups with 30 speakers each, covering pediatric and adult populations and resonance-, motor-, and laryngeal-based pathologies. The 10-listener panel crosses two attributes, native language and domain expertise, whose effects on perception are examined in subsequent sections. Domain expertise is defined as at least 8 years of specialized experience in signal-level speech processing or in clinical voice and speech assessment. Ages are reported as mean $\pm$ standard deviation (SD) [range] in years.}
\label{tab:cohort}
\setlength{\tabcolsep}{4pt}
\renewcommand{\arraystretch}{1.08}
\begin{tabular}{@{}lccl@{}}
\toprule
\multicolumn{4}{l}{\textit{Speaker cohort}} \\
\midrule
Group & $n$ & Female / Male & Age (yr), mean $\pm$ SD [range] \\
\midrule
CLP            & 30  & 19 / 11           & 12 $\pm$ \phantom{0}3 \phantom{0}[6, 18]  \\
Dysarthria     & 30  & 13 / 17           & 50 $\pm$ 18 [20, 75]                       \\
Dysglossia     & 30  & 16 / 14           & 58 $\pm$ 18 [24, 78]                       \\
Dysphonia      & 30  & \phantom{0}5 / 25 & 59 $\pm$ 12 [24, 76]                       \\
Adult controls & 30  & 20 / 10           & 19 $\pm$ \phantom{0}7 [11, 37]             \\
Child controls & 30  & 20 / 10           & 11 $\pm$ \phantom{0}3 \phantom{0}[7, 16]   \\
\midrule
Total          & 180 & 93 / 87           & 35 $\pm$ 24 \phantom{0}[6, 78]             \\
\midrule
\multicolumn{4}{l}{\textit{Listener panel}} \\
\midrule
Listener & Native language & German proficiency & Domain expertise \\
\midrule
L1  & Persian  & A1     & Non-expert                              \\
L2  & Spanish  & B2     & Expert (speech processing)              \\
L3  & Mandarin & C1     & Expert (speech processing)              \\
L4  & Persian  & B1     & Non-expert                              \\
L5  & Persian  & B1     & Non-expert                              \\
L6  & German   & Native & Non-expert                              \\
L7  & German   & Native & Non-expert (clinical radiologist)       \\
L8  & German   & Native & Expert (clinical phoniatrician)         \\
L9  & German   & Native & Non-expert                              \\
L10 & German   & Native & Expert (speech processing)              \\
\bottomrule
\end{tabular}
\end{table}


\subsection*{Anonymization is perceptually detectable with pathology-dependent salience}

\begin{figure}[p]
\centering
\includegraphics[width=\textwidth]{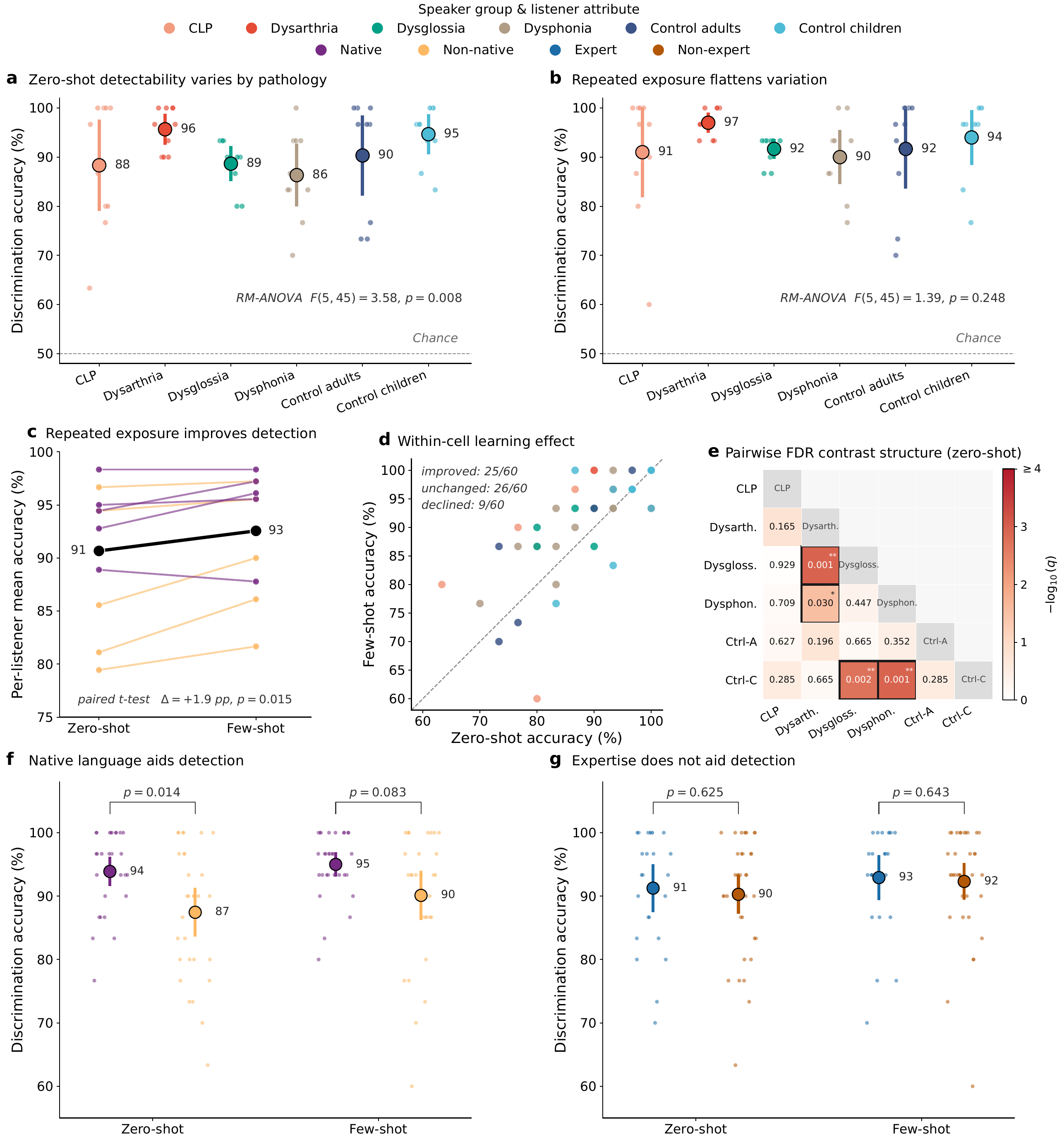}
\caption{Anonymization is perceptually detectable with pathology-dependent salience.
\textbf{a},\textbf{b} Discrimination accuracy across the six speaker groups in the zero-shot (\textbf{a}) and few-shot (\textbf{b}) Turing-style tasks. Small dots are individual listeners ($n=10$), the vertical bar is the 95\% CI of the group mean, and the large dot marks the mean. The dashed line marks 50\% chance.
\textbf{c} Per-listener mean accuracy from zero-shot to few-shot. Each line is one listener, colored by native-language status (see top legend); the black trace is the cohort mean.
\textbf{d} Within-cell learning: each point is one (listener, group) cell ($n=60$); $x$-axis zero-shot, $y$-axis few-shot; dashed line $y=x$.
\textbf{e} Lower-triangular matrix of FDR-corrected $q$-values for all 15 pairwise contrasts on zero-shot accuracy (paired $t$-test). Cell color encodes $-\log_{10}(q)$; cells with $q<0.05$ are outlined in black and annotated with asterisks ($*$ $q<0.05$, $**$ $q<0.01$, $***$ $q<0.001$).
\textbf{f},\textbf{g} Cell-level discrimination accuracy stratified by listener native language (\textbf{f}; native $n=30$, non-native $n=30$) and domain expertise (\textbf{g}; expert $n=24$, non-expert $n=36$). Small dots are individual cell-level values, vertical bars are 95\% CIs, large dots are means. Two-sided Mann-Whitney $U$ $p$-values shown above brackets.}
\label{fig:discrimination}
\end{figure}

Listeners discriminated the anonymized from the original recordings at 91 $\pm$ 7 [86, 95] zero-shot and 93 $\pm$ 6 [88, 97] few-shot accuracy (per-listener means, $n=10$), both far above the 50 chance level (Fig.~\ref{fig:discrimination}a,b). The automatic transformation is therefore not perceptually transparent in clinical recordings: even on first exposure and without any familiarization, a listener can identify the processed file roughly nine times out of ten. The salience of the transformation, however, was not uniform across pathologies. Zero-shot accuracy differed significantly across the six groups (repeated-measures one-way ANOVA \cite{OneWayANOVA,BasicandAdvancedStatisticalTests,anova_main,repeatedmeasures}, $F(5,45) = 3.579$, $p = 0.008$; all statistical tests reported are tabulated in Supplementary Table~\ref{stab:all_tests}), with Dysarthria (96 $\pm$ 4 [92, 99]) and child controls (95 $\pm$ 6 [91, 99]) at the high end and Dysphonia (86 $\pm$ 9 [80, 93]) at the low end. Pairwise contrasts with Benjamini-Hochberg false discovery rate (FDR) correction \cite{benjamini1995controlling} isolated a specific structure rather than a diffuse spread: only four of the fifteen group pairs reached significance, namely Dysarthria $>$ Dysglossia ($q = 0.001$), Dysarthria $>$ Dysphonia ($q = 0.030$), child controls $>$ Dysglossia ($q = 0.002$), and child controls $>$ Dysphonia ($q = 0.001$), with Dysphonia and Dysglossia as the consistently less detectable member of every significant pair (Fig.~\ref{fig:discrimination}e). The fact that detectability tracks pathology, rather than being uniform across groups, suggests that residual disorder-correlated acoustic cues survive the McAdams formant shift and that these residual cues are differentially accessible to human listeners depending on the underlying disorder.

A brief familiarization phase narrowed this pathology-driven variation. In the few-shot task the across-group difference no longer reached significance ($F(5,45) = 1.386$, $p = 0.248$; Fig.~\ref{fig:discrimination}b), while overall accuracy increased modestly but reliably ($\Delta = +1.9$ percentage points, paired two-tailed t-test, $p = 0.015$; Fig.~\ref{fig:discrimination}c). At the listener-by-group cell level, 25 of 60 cells improved with familiarization, 26 were unchanged, and 9 declined (Fig.~\ref{fig:discrimination}d). The cue underlying anonymization detection is therefore largely accessible at first exposure rather than emerging only after training; familiarization smooths inter-pathology differences instead of uncovering a qualitatively new signal.

Two listener attributes were examined for modulation of the detection signal. Native German speakers outperformed non-native listeners in the zero-shot task (94 $\pm$ 6 [92, 96] vs 87 $\pm$ 10 [84, 91], two-sided Mann-Whitney U \cite{mann1947test,mannwit} on the 30 cell-level accuracies per subgroup, $p = 0.014$), with the difference attenuating in the few-shot task (95 $\pm$ 5 [93, 97] vs 90 $\pm$ 10 [86, 94], $p = 0.083$; Fig.~\ref{fig:discrimination}f). Domain expertise, by contrast, produced no detectable effect in either task: experts and non-experts performed within a single percentage point of each other (zero-shot 91 $\pm$ 9 [87, 95] vs 90 $\pm$ 9 [87, 93], $p = 0.625$; few-shot 93 $\pm$ 8 [89, 96] vs 92 $\pm$ 9 [89, 95], $p = 0.643$; Fig.~\ref{fig:discrimination}g). These results locate the discrimination signal in low-level acoustic-perceptual familiarity rather than in clinical or signal-processing training: the cue is salient to anyone who has spoken German since childhood, irrespective of professional background.

\subsection*{Anonymization reduces perceived speech quality with disorder-specific degradation}

\begin{figure}[p]
\centering
\includegraphics[width=\textwidth]{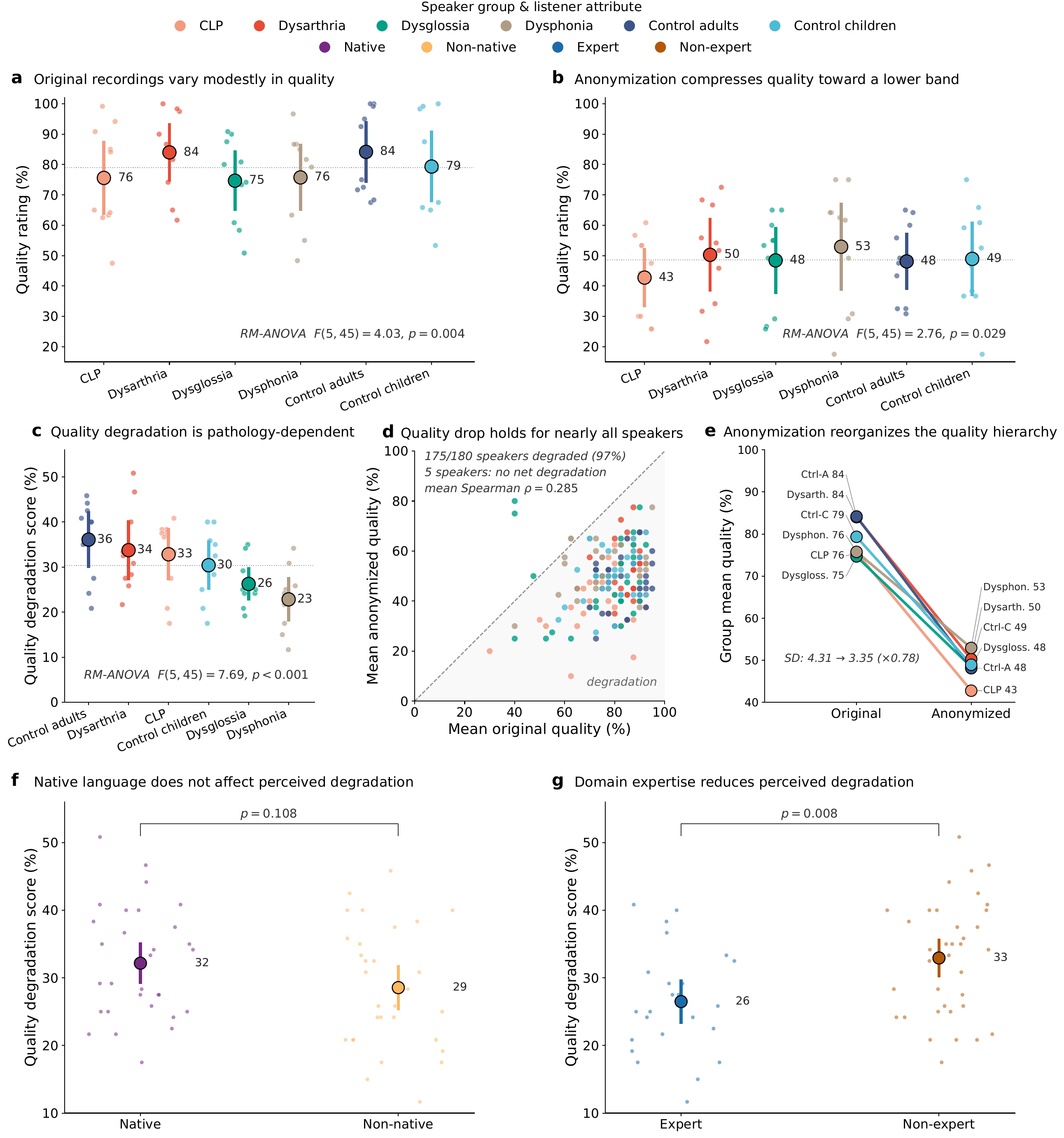}
\caption{Anonymization reduces perceived speech quality with disorder-specific degradation.
\textbf{a},\textbf{b} Per-listener mean quality rating across the six speaker groups for the original (\textbf{a}) and anonymized (\textbf{b}) recordings, with the native 5-point scale linearly mapped to 0--100. Small dots are individual listeners ($n=10$), the vertical bar is the 95\% CI of the group mean, the large dot marks the mean, and the dotted horizontal line marks the cohort mean. Repeated-measures one-way ANOVA across the six groups.
\textbf{c} Quality degradation score (QDS, original rating minus anonymized rating) by group, ordered by descending mean; same plotting convention as \textbf{a},\textbf{b}, with the dotted line marking the overall mean QDS.
\textbf{d} Per-speaker mean original vs mean anonymized quality, averaged across the 10 listeners; one dot per speaker ($n=180$), colored by group, dashed line $y=x$. Points below the diagonal indicate net degradation. Inter-listener agreement on per-speaker QDS is the average pairwise Spearman correlation.
\textbf{e} Slopegraph of each group's mean quality from the original (left axis) to the anonymized (right axis) condition; line crossings indicate rank reorganization of the perceived-quality hierarchy.
\textbf{f},\textbf{g} Cell-level QDS stratified by listener native language (\textbf{f}; native $n=30$, non-native $n=30$) and domain expertise (\textbf{g}; expert $n=24$, non-expert $n=36$). Small dots are individual cell-level values, vertical bars are 95\% CIs, large dots are means; two-sided Mann-Whitney $U$ $p$-values above brackets.}
\label{fig:quality}
\end{figure}

Beyond detectability, the automatic transformation produced a substantial drop in perceived speech quality. With the 5-point quality rating linearly mapped to a 0--100 scale ($1 \rightarrow 0$, $5 \rightarrow 100$), original recordings were rated at 79 $\pm$ 14 [69, 89] and anonymized recordings at 49 $\pm$ 15 [38, 59] (per-listener means, $n=10$; paired two-tailed t-test, $p<0.001$; Fig.~\ref{fig:quality}a,b). The quality degradation score (QDS), defined per (listener, speaker) pair as the original rating minus the anonymized rating, averaged 30 $\pm$ 6 [26, 35] percentage points. The drop was nearly universal at the speaker level: 175 of 180 speakers (97\%) received a positive mean QDS across listeners, with only 5 showing no net degradation (Fig.~\ref{fig:quality}d). The McAdams shift is therefore not merely detectable: it is perceived as a tangible reduction in vocal naturalness in essentially every speaker. The full $10 \times 6$ per-listener per-group accuracy and QDS matrices that underlie these and the preceding analyses are provided in Supplementary Table~\ref{stab:cell_matrices}.

The magnitude of degradation, however, depended strongly on the underlying pathology (repeated-measures one-way ANOVA on per-listener QDS, $F(5,45)=7.690$, $p<0.001$). The largest drops occurred in adult controls (36 $\pm$ 9 [30, 42]), Dysarthria (34 $\pm$ 9 [27, 40]), and CLP (33 $\pm$ 8 [27, 39]), while Dysphonia (23 $\pm$ 7 [18, 28]) and Dysglossia (26 $\pm$ 5 [23, 30]) showed the smallest drops (Fig.~\ref{fig:quality}c). After FDR correction, seven pairwise contrasts reached significance, all consistent with the same gradient: Dysphonia and Dysglossia degraded less than CLP, Dysarthria, and the control groups ($q \leq 0.046$ for all). The asymmetry produces a marked reorganization of the perceived-quality hierarchy (Fig.~\ref{fig:quality}e). Original-quality means span Dysarthria and the adult controls at the top (both 84) and Dysglossia at the bottom (75), but after anonymization Dysphonia ends with the highest mean rating (53) and CLP the lowest (43). The standard deviation across the six group means falls from 4.31 in the original condition to 3.35 in the anonymized condition (ratio 0.78), indicating partial convergence of inter-group quality differences once the anonymization signature is overlaid. A parsimonious interpretation is a perceptual floor effect: pathologies whose original recordings already sit lower on the naturalness axis have less headroom to lose, and the anonymization shift is rated more leniently relative to a degraded baseline. The complete pairwise table including non-significant contrasts is given in Supplementary Table~\ref{stab:pairwise_full}.

Listener attributes affected the QDS asymmetrically with respect to the discrimination findings. Native and non-native listeners produced similar QDS values (32 $\pm$ 6 [25, 40] vs 29 $\pm$ 6 [22, 36], two-sided Mann-Whitney U on the 30 cell-level QDS values per subgroup, $p=0.108$; Fig.~\ref{fig:quality}f). Experts and non-experts, in contrast, differed substantially: experts reported a markedly smaller QDS than non-experts (26 $\pm$ 6 [17, 36] vs 33 $\pm$ 5 [28, 38], $p=0.008$; Fig.~\ref{fig:quality}g). This pattern is the inverse of the discrimination task, in which expertise was uninformative but native language mattered, and indicates that domain training calibrates the magnitude of perceived degradation even though it does not improve the ability to detect that degradation has occurred. Inter-listener agreement on which speakers were most degraded was modest (mean Spearman $\rho$ across listener pairs on per-speaker QDS = 0.285, range [0.009, 0.497]; Fig.~\ref{fig:quality}d), consistent with substantial idiosyncratic variation in how individual listeners weight the anonymization-induced spectral change.

\subsection*{Clinical pathology severity is preserved after anonymization}

The perceptual findings of the preceding sections leave open the clinically central question: does the automatic transformation preserve the diagnostic signal? Detectability and a quality drop matter for downstream listener experience, but a release of anonymized recordings for clinical purposes hinges on whether the same clinician would arrive at the same severity judgment on the anonymized file. To address this directly, the senior phoniatrician on the listening panel (M.S. with 35 years of experience) rated all 360 recordings under blinding, on the standard four-point ordinal scale used in clinical voice and speech assessment (0 none, 1 mild, 2 moderate, 3 severe). Files were presented in randomized order with the original-versus-anonymized position concealed.

\begin{figure}[p]
\centering
\includegraphics[width=\textwidth]{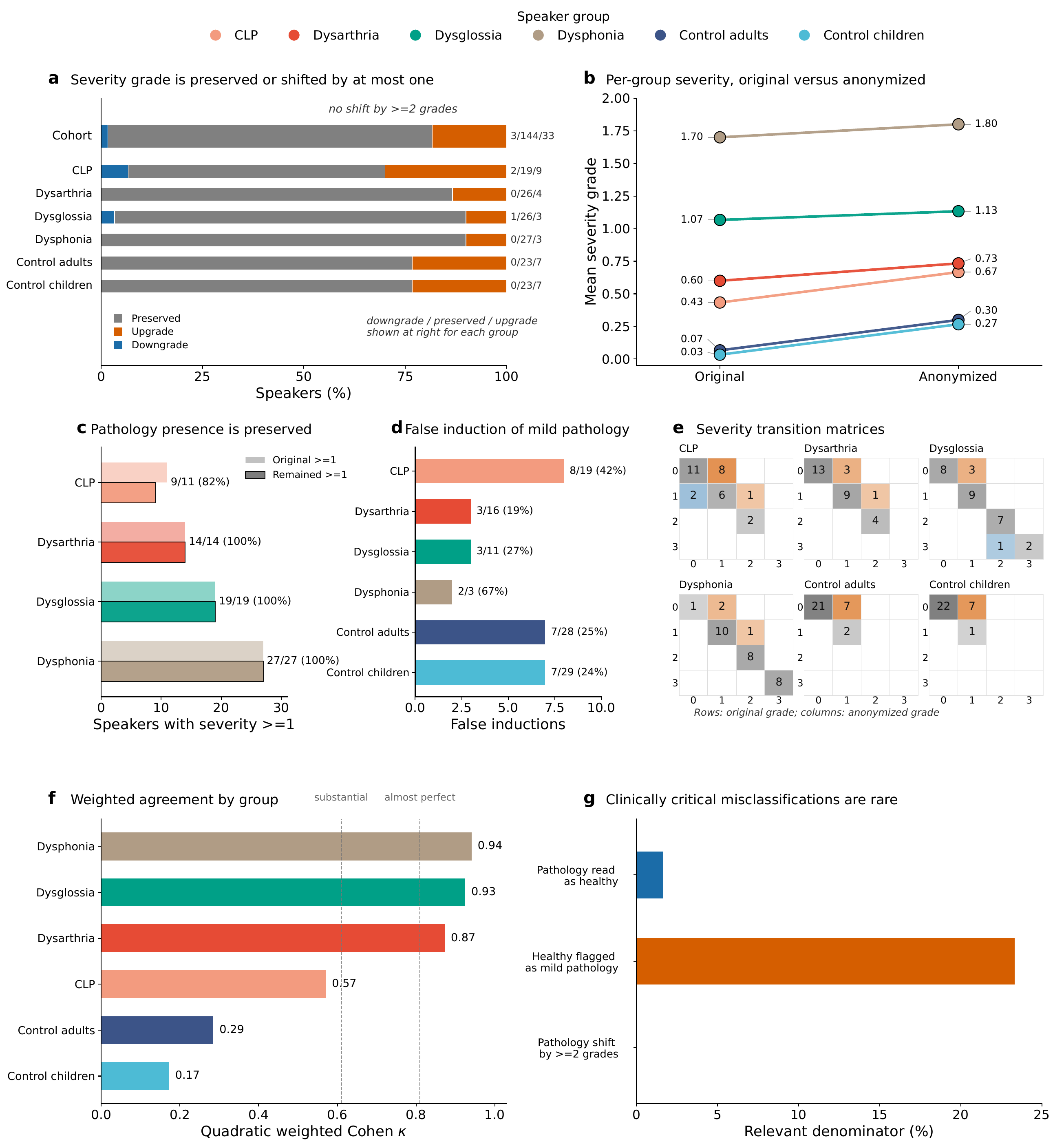}
\caption{Clinical pathology severity is preserved after anonymization.
\textbf{a} Cohort-level distribution of the within-speaker severity shift $\Delta = $ anonymized $-$ original (top), with per-group breakdown of the same shift composition (bottom). Bars are color-coded by direction: preserved (gray, $\Delta = 0$), upgrade (red-orange, $\Delta = +1$), downgrade (blue, $\Delta = -1$).
\textbf{b} Slopegraph of each group's mean severity from the original (left axis) to the anonymized (right axis) condition.
\textbf{c} Pathology-presence preservation in the four pathology groups: for each group, the number of speakers originally rated as having any pathology (severity $\geq 1$, upper bars) and the number who remained at $\geq 1$ after anonymization (lower bars, outlined).
\textbf{d} False induction across all six groups: number of originally-grade-0 recordings whose anonymized version was upgraded to grade $\geq 1$.
\textbf{e} Per-group $4 \times 4$ severity transition matrices (rows: original grade, columns: anonymized grade). Diagonal cells (preserved) are gray, above-diagonal cells (upgrade) red-orange, below-diagonal cells (downgrade) blue; color intensity scales with count and cell values are printed inside.
\textbf{f} Quadratic-weighted Cohen's $\kappa$ by group, sorted descending. Vertical dashed lines mark the conventional boundaries at $\kappa = 0.61$ (substantial) and $\kappa = 0.81$ (almost perfect).
\textbf{g} Directional summary of clinically relevant misclassifications. Each bar is normalized to its own denominator.}
\label{fig:severity}
\end{figure}

Across the full cohort, severity grades remained close. Of the 180 speakers, 144 (80\%) received the identical grade in both conditions and the remaining 36 shifted by exactly one grade; no speaker shifted by two or more grades in either direction (Fig.~\ref{fig:severity}a). Among the 36 shifts, 33 were upward and only 3 were downward, indicating a small but systematic over-call rather than random noise. The per-group decomposition in the lower half of Fig.~\ref{fig:severity}a shows the upward shifts concentrating in CLP and the two control groups, with the three adult pathology groups (Dysarthria, Dysglossia, Dysphonia) closer to full preservation. The same picture emerges at the group-mean level: every group's mean severity rose only marginally between original and anonymized recordings, with the cohort Wilcoxon signed-rank test yielding $p < 0.001$ purely on the strength of the consistent upward direction rather than on shift magnitude (Fig.~\ref{fig:severity}b).

At the pathology level the diagnostic signal survives almost intact (Fig.~\ref{fig:severity}c and Table~\ref{tab:severity}). Among the 71 pathology speakers (CLP, Dysarthria, Dysglossia, Dysphonia) originally rated as having any pathology (severity $\geq 1$), 69 (97\%) remained at grade 1 or higher after anonymization; the two exceptions were both CLP speakers who slipped from grade 1 to grade 0. The per-group transition matrices (Fig.~\ref{fig:severity}e) localize the residual shifts: in Dysarthria, Dysglossia, and Dysphonia, almost all shifts that did occur were single-step movements adjacent to the diagonal (e.g. 0 to 1, or 1 to 2), and the quadratic-weighted Cohen's $\kappa$ \cite{cohen1968weighted} for these three groups was 0.87, 0.93, and 0.94, respectively, all in the conventional "almost perfect" range (Fig.~\ref{fig:severity}f). CLP was the only pathology group with lower agreement (exact 63\%, $\kappa_w = 0.57$), driven mostly by 8 of 19 originally-grade-0 CLP recordings being upgraded to grade 1 after anonymization (Fig.~\ref{fig:severity}d); the within-group Wilcoxon test indicated a small upward shift that did not survive FDR correction across the six groups (raw $p = 0.035$, $q = 0.068$). Per-group Wilcoxon $p$-values for the original-versus-anonymized comparison, FDR-corrected across the six groups \cite{benjamini1995controlling}, are reported in Table~\ref{tab:severity}.

The upward bias was concentrated, as expected, in the controls. Of the 60 control speakers, 14 (23\%) originally rated as having no pathology were newly flagged at grade 1 after anonymization (Fig.~\ref{fig:severity}d), and the within-group Wilcoxon tests for adult and child controls were the only two of the six contrasts to survive FDR correction (raw $p = 0.008$, $q = 0.024$ each). Crucially, no control was upgraded beyond mild severity. The clinically dangerous direction, an anonymized pathological recording being read as healthy, is therefore very rare (2 of 120 pathology speakers, both mild-grade CLP), and severity escalation beyond mild within the pathology cohort is rare and never crossed the mild-to-moderate boundary for more than 3 of the 39 originally-mild pathology speakers (Fig.~\ref{fig:severity}g). The more common misreading is a healthy recording acquiring a mild-pathology flag because the residual McAdams noise is read as hoarseness, an interpretive caveat that the rating clinician explicitly raised and that a clinically-aware reader can correct for. Taken together with the perceptual conspicuousness documented in the preceding sections, the severity results indicate that the automatic transformation, despite being perceptually salient and reducing perceived speech quality, leaves the diagnostic pathology signal substantively intact at the pathology level and within each clinical group.

\begin{table}[p]
\centering
\caption{Pathology severity preservation under McAdams anonymization, rated by a senior phoniatrician (L8) under blinding on the 0--3 ordinal scale. $\Delta$severity is anonymized minus original; positive values indicate over-call. Pathology-preservation rate is the fraction of speakers originally rated as having pathology (severity $\geq 1$) who remained at grade $\geq 1$ after anonymization; this construct is undefined for control groups, where false-induction rate is reported instead. Exact-grade agreement is the fraction of speakers receiving identical grades in both conditions; quadratic-weighted Cohen's $\kappa$ summarizes ordinal agreement. Wilcoxon $p$-values come from per-group signed-rank tests of original vs anonymized severity, with $q$-values reflecting FDR correction across the six groups. N/A, not assigned.}
\label{tab:severity}
\setlength{\tabcolsep}{4pt}
\renewcommand{\arraystretch}{1.05}
\begin{tabular}{@{}p{0.21\columnwidth}p{0.2\columnwidth}p{0.3\columnwidth}p{0.24\columnwidth}@{}}
\toprule
Group & Severity ratings & Preservation / induction & Agreement and tests \\
\midrule
\multicolumn{4}{@{}l}{\textit{Pathology groups}} \\
\midrule
CLP, $n=30$ &
\begin{tabular}[t]{@{}ll@{}}
Orig. & 0.43 $\pm$ 0.63 \\
Anon. & 0.67 $\pm$ 0.66 \\
$\Delta$ & $+$0.23
\end{tabular}
&
\begin{tabular}[t]{@{}ll@{}}
Pathology pres. & 9/11 (82\%) \\
False induc. & 8/19 (42\%)
\end{tabular}
&
\begin{tabular}[t]{@{}ll@{}}
Exact agree. & 19/30 (63\%) \\
$\kappa_w$ & 0.57 \\
$p$ & 0.035 \\
$q$ & 0.068
\end{tabular}
\\
\midrule
Dysarthria, $n=30$ &
\begin{tabular}[t]{@{}ll@{}}
Orig. & 0.60 $\pm$ 0.72 \\
Anon. & 0.73 $\pm$ 0.74 \\
$\Delta$ & $+$0.13
\end{tabular}
&
\begin{tabular}[t]{@{}ll@{}}
Pathology pres. & 14/14 (100\%) \\
False induc. & 3/16 (19\%)
\end{tabular}
&
\begin{tabular}[t]{@{}ll@{}}
Exact agree. & 26/30 (87\%) \\
$\kappa_w$ & 0.87 \\
$p$ & 0.046 \\
$q$ & 0.068
\end{tabular}
\\
\midrule
Dysglossia, $n=30$ &
\begin{tabular}[t]{@{}ll@{}}
Orig. & 1.07 $\pm$ 1.01 \\
Anon. & 1.13 $\pm$ 0.90 \\
$\Delta$ & $+$0.07
\end{tabular}
&
\begin{tabular}[t]{@{}ll@{}}
Pathology pres. & 19/19 (100\%) \\
False induc. & 3/11 (27\%)
\end{tabular}
&
\begin{tabular}[t]{@{}ll@{}}
Exact agree. & 26/30 (87\%) \\
$\kappa_w$ & 0.93 \\
$p$ & 0.317 \\
$q$ & 0.317
\end{tabular}
\\
\midrule
Dysphonia, $n=30$ &
\begin{tabular}[t]{@{}ll@{}}
Orig. & 1.70 $\pm$ 0.99 \\
Anon. & 1.80 $\pm$ 0.89 \\
$\Delta$ & $+$0.10
\end{tabular}
&
\begin{tabular}[t]{@{}ll@{}}
Pathology pres. & 27/27 (100\%) \\
False induc. & 2/3 (67\%)
\end{tabular}
&
\begin{tabular}[t]{@{}ll@{}}
Exact agree. & 27/30 (90\%) \\
$\kappa_w$ & 0.94 \\
$p$ & 0.083 \\
$q$ & 0.100
\end{tabular}
\\
\midrule
\multicolumn{4}{@{}l}{\textit{Control groups}} \\
\midrule
Adult controls, $n=30$ &
\begin{tabular}[t]{@{}ll@{}}
Orig. & 0.07 $\pm$ 0.25 \\
Anon. & 0.30 $\pm$ 0.47 \\
$\Delta$ & $+$0.23
\end{tabular}
&
\begin{tabular}[t]{@{}ll@{}}
Pathology pres. & N/A \\
False induc. & 7/28 (25\%)
\end{tabular}
&
\begin{tabular}[t]{@{}ll@{}}
Exact agree. & 23/30 (77\%) \\
$\kappa_w$ & 0.29 \\
$p$ & 0.008 \\
$q$ & 0.024
\end{tabular}
\\
\midrule
Child controls, $n=30$ &
\begin{tabular}[t]{@{}ll@{}}
Orig. & 0.03 $\pm$ 0.18 \\
Anon. & 0.27 $\pm$ 0.45 \\
$\Delta$ & $+$0.23
\end{tabular}
&
\begin{tabular}[t]{@{}ll@{}}
Pathology pres. & N/A \\
False induc. & 7/29 (24\%)
\end{tabular}
&
\begin{tabular}[t]{@{}ll@{}}
Exact agree. & 23/30 (77\%) \\
$\kappa_w$ & 0.17 \\
$p$ & 0.008 \\
$q$ & 0.024
\end{tabular}
\\
\bottomrule
\end{tabular}
\end{table}

\subsection*{Listener language and expertise modulate perception; speaker demographics do not}

\begin{figure}[p]
\centering
\includegraphics[width=\textwidth]{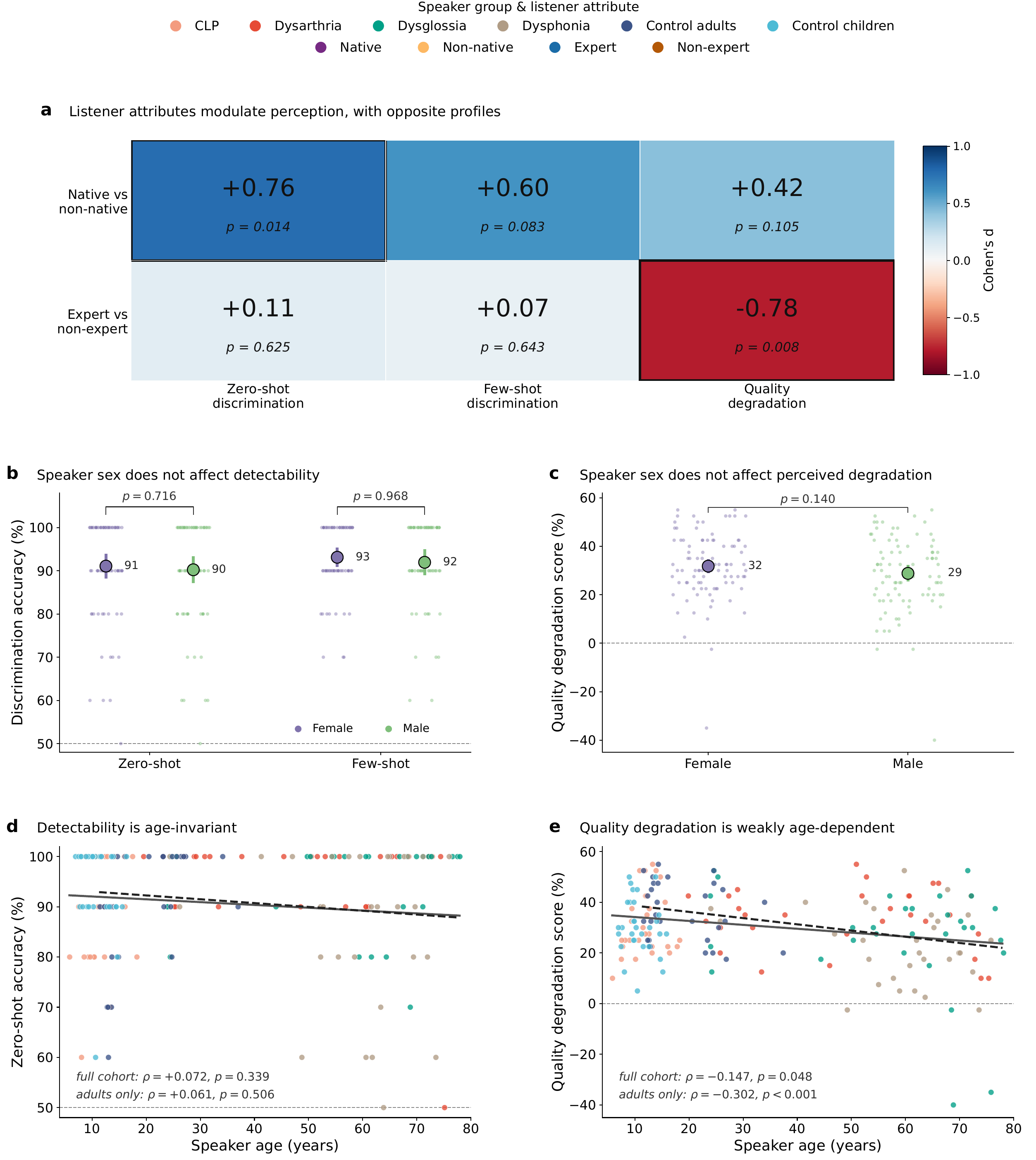}
\caption{Listener language and expertise modulate perception; speaker demographics do not.
\textbf{a} Cohen's $d$ effect sizes for the two listener-attribute contrasts (rows: native vs non-native; expert vs non-expert) across the three perceptual outcomes (columns: zero-shot discrimination, few-shot discrimination, quality degradation score). Positive $d$ favors the native or expert group; cell color encodes $d$ on the diverging scale at right, and cells with a two-sided Mann-Whitney $U$ $p<0.05$ are outlined in black. The native-language effect is confined to discrimination, whereas the expertise effect is confined to quality degradation.
\textbf{b},\textbf{c} Per-speaker discrimination accuracy (\textbf{b}; zero-shot and few-shot) and QDS (\textbf{c}) stratified by speaker sex; small dots are individual speakers averaged across the 10 listeners, vertical bars are 95\% CIs, large dots are means. Two-sided Mann-Whitney $U$ $p$-values are shown above brackets.
\textbf{d},\textbf{e} Per-speaker zero-shot accuracy (\textbf{d}) and QDS (\textbf{e}) as a function of speaker age, one dot per speaker colored by group. The solid line is the full-cohort linear fit, the dashed line the adult-only fit ($n=120$); Spearman $\rho$ and $p$ for both are annotated. Few-shot accuracy shows the same age-invariance as \textbf{e} and is reported in the text.}
\label{fig:demographics}
\end{figure}

Placing the two listener attributes side by side across all three perceptual outcomes exposes a clean double dissociation (Fig.~\ref{fig:demographics}a). The native-language advantage is confined to the discrimination task, where it is moderate-to-large (Cohen's \cite{cohen2013statistical} $d=0.76$ zero-shot, $d=0.60$ few-shot), and falls to a small, non-significant effect on quality degradation ($d=0.42$, two-sided Mann-Whitney U $p=0.108$). Domain expertise shows the mirror-image profile: negligible on discrimination ($d=0.11$ zero-shot, $d=0.07$ few-shot) but large on perceived degradation, with experts reporting a substantially smaller QDS than non-experts ($d=-0.78$, $p=0.008$). Detecting the anonymization signature therefore draws on perceptual familiarity with the native language, while the magnitude of perceived quality loss is calibrated by clinical or signal-processing training. Because neither attribute covers both axes, a deployment evaluation that recruits only one type of listener will systematically misjudge one of the two outcomes.

Speaker sex, by contrast, produced no detectable bias on any outcome. Female and male speakers were discriminated at statistically indistinguishable rates (zero-shot 91 $\pm$ 14 [88, 94] vs 90 $\pm$ 15 [87, 93], two-sided Mann-Whitney U on per-speaker values, $p=0.716$; few-shot 93 $\pm$ 11 [91, 95] vs 92 $\pm$ 14 [89, 95], $p=0.968$; Fig.~\ref{fig:demographics}b) and received comparable QDS (32 $\pm$ 14 [29, 35] vs 29 $\pm$ 16 [25, 32], $p=0.140$; Fig.~\ref{fig:demographics}c). This fairness held inside every pathology group: across all 18 within-group sex contrasts spanning the three outcomes, none reached significance, the smallest $p$-value being 0.097. The result is reassuring even for Dysphonia, whose strongly male-skewed composition (25 male, 5 female) could in principle have masked or exaggerated a disparity. Per-group female-versus-male breakdowns are provided in Supplementary Table~\ref{stab:sex_contrasts}. The McAdams transformation thus did not advantage one sex over the other, a property that matters for clinical deployment in pathology cohorts whose sex distribution is often uneven.

Speaker age had at most a weak influence, and only on perceived quality. Age was uncorrelated with discrimination accuracy across the full cohort (Spearman $\rho=+0.072$, $p=0.339$ zero-shot; $\rho=+0.138$, $p=0.065$ few-shot; Fig.~\ref{fig:demographics}d) and showed only a shallow negative association with QDS ($\rho=-0.147$, $p=0.048$; Fig.~\ref{fig:demographics}e). Restricting to the four adult groups ($n=120$) left the discrimination relationship negligible (zero-shot $\rho=+0.061$, $p=0.506$; few-shot $\rho=+0.064$, $p=0.487$) but modestly strengthened the age-QDS association ($\rho=-0.302$, $p=0.001$). Older speakers therefore tended to receive slightly smaller quality penalties under anonymization, plausibly a within-cohort echo of the floor effect seen across pathologies (Fig.~\ref{fig:quality}c): voices that already sit lower on the naturalness axis have less headroom to lose. The detectability of the transformation itself, however, is effectively age-invariant.

\subsection*{Perceptual outcomes are decoupled from computational privacy metrics}

\begin{figure}[p]
\centering
\includegraphics[width=\textwidth]{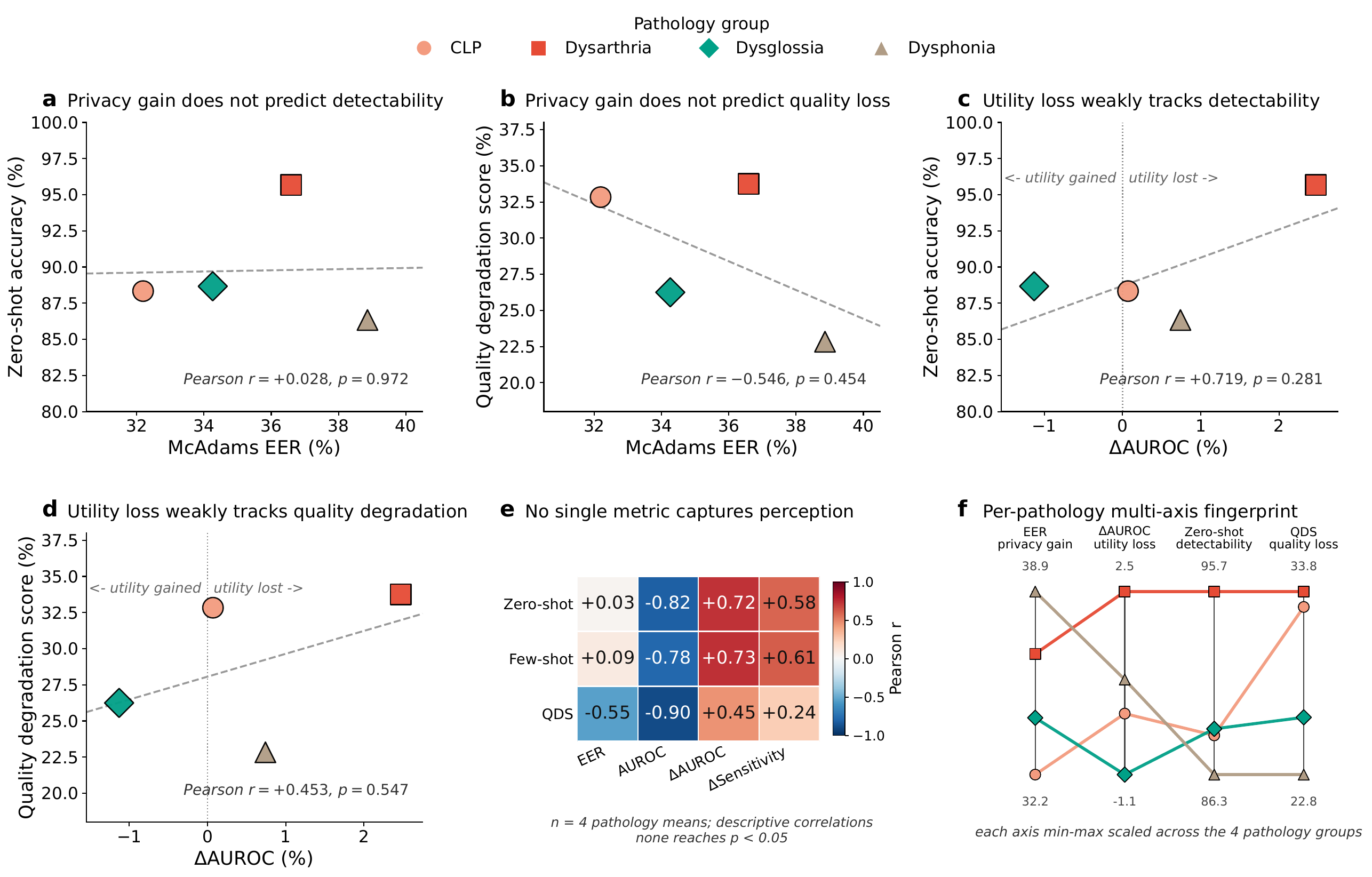}
\caption{Perceptual outcomes are decoupled from computational privacy and only weakly track utility metrics.
All panels use the four pathology groups (CLP, Dysarthria, Dysglossia, Dysphonia); controls are excluded because the downstream pathology classifier defines them only as the negative class. Computational values (EER, AUROC, $\Delta\text{AUROC}$, $\Delta$sensitivity) are taken from that prior work; perceptual values are computed here.
\textbf{a},\textbf{b} Per-group zero-shot discrimination accuracy (\textbf{a}) and QDS (\textbf{b}) as a function of EER under McAdams anonymization. Each dot is one pathology group; the dashed line is the least-squares trend.
\textbf{c},\textbf{d} Zero-shot discrimination (\textbf{c}) and QDS (\textbf{d}) as a function of utility loss $\Delta\text{AUROC}$ (original minus anonymized; positive values indicate utility decreases after anonymization). The vertical reference line marks $\Delta\text{AUROC}=0$.
\textbf{e} Pearson correlation matrix between the three perceptual measures (rows) and four computational measures (columns); cell color encodes $r$. With $n=4$ pathology means the correlations are descriptive and none reach $p<0.05$.
\textbf{f} Per-pathology profile across four axes (EER, $\Delta\text{AUROC}$, zero-shot detectability, QDS), each min--max scaled across the four groups; each line is one pathology. Crossing lines indicate that privacy gain, utility loss, and perceptual cost do not co-order across pathologies.}
\label{fig:disconnect}
\end{figure}

The McAdams anonymization has previously been characterized on the same speaker pool through two computational summaries: the EER of an ASV system, where higher values indicate stronger anonymization, and the AUROC of a downstream pathology classifier, where the drop from original to anonymized indexes utility loss \cite{arasteh2024addressing}. Whether either metric anticipates the perceptual outcomes documented above is unknown, and the question matters in practice, since deployment decisions are typically grounded on computational evidence alone. For the four pathology groups with paired computational measurements (CLP, Dysarthria, Dysglossia, Dysphonia), the McAdams EER, post-anonymization AUROC, and AUROC change ($\Delta\text{AUROC}$ = original $-$ anonymized; positive values indicate utility loss) were placed alongside the per-group perceptual measures (Fig.~\ref{fig:disconnect}).

Privacy gain and perceptual salience were essentially unrelated across the four groups. EER did not track zero-shot detectability (Pearson $r = +0.028$, $p = 0.972$; $n = 4$ pathology means; Fig.~\ref{fig:disconnect}a), few-shot detectability ($r = +0.092$, $p = 0.908$), or quality degradation ($r = -0.546$, $p = 0.454$; Fig.~\ref{fig:disconnect}b). The dissociation is sharpest at the two extremes. Dysphonia carries the highest McAdams EER of the four groups (39\%, the strongest computational anonymization) yet the lowest zero-shot detectability (86) and the smallest QDS (23) of any group in the cohort. CLP shows the inverse arrangement: the lowest EER (32\%, the weakest computational anonymization here) but mid-range detectability (88) and a comparatively large QDS (33). An evaluator relying on EER alone would rank Dysphonia the cleanest anonymization and CLP the weakest, whereas the perceptual evidence reverses that ordering on QDS and places the two groups on essentially equal footing on detectability. EER and perception therefore index different facets of the transformation: speaker-identity confusability on one hand, naturalness change on the other.

Utility change carried a weak perceptual signature, consistent in sign but well short of significance at the four available group means. Post-anonymization AUROC was inversely associated with zero-shot detectability ($r = -0.817$, $p = 0.183$) and with QDS ($r = -0.897$, $p = 0.103$), and $\Delta\text{AUROC}$ was positively associated with detectability ($r = +0.719$, $p = 0.281$; Fig.~\ref{fig:disconnect}c) and with QDS ($r = +0.453$, $p = 0.547$; Fig.~\ref{fig:disconnect}d). The direction is interpretable: Dysarthria loses the most utility under anonymization ($\Delta\text{AUROC} = 2.47$) and is simultaneously the most detectable group with the largest QDS, while Dysglossia, whose downstream AUROC marginally improves after anonymization ($\Delta\text{AUROC} = -1.13$), sits in the middle of the perceptual distributions. The correlation matrix across three perceptual and four computational measures shows the same picture: every cell is directional but none reaches significance at $n = 4$ (Fig.~\ref{fig:disconnect}e). No single computational metric individually predicts the perceptual response with the precision a deployment evaluation would require.

The consequence is visible in the per-pathology profiles (Fig.~\ref{fig:disconnect}f). A pipeline validated purely on EER and AUROC, as is standard in the speaker-anonymization literature, would judge all four pathologies broadly successful applications of the McAdams method; the perceptual evidence shows instead that the transformation is highly conspicuous in Dysarthria, moderately so in CLP, and comparatively unobtrusive in Dysphonia and Dysglossia. Computational and perceptual evaluations capture complementary, non-redundant information, and a clinically responsible release of anonymized speech should report both.

\subsection*{Disorder-specific perceptual signatures emerge across all measures}

\begin{figure}[p]
\centering
\includegraphics[width=\textwidth]{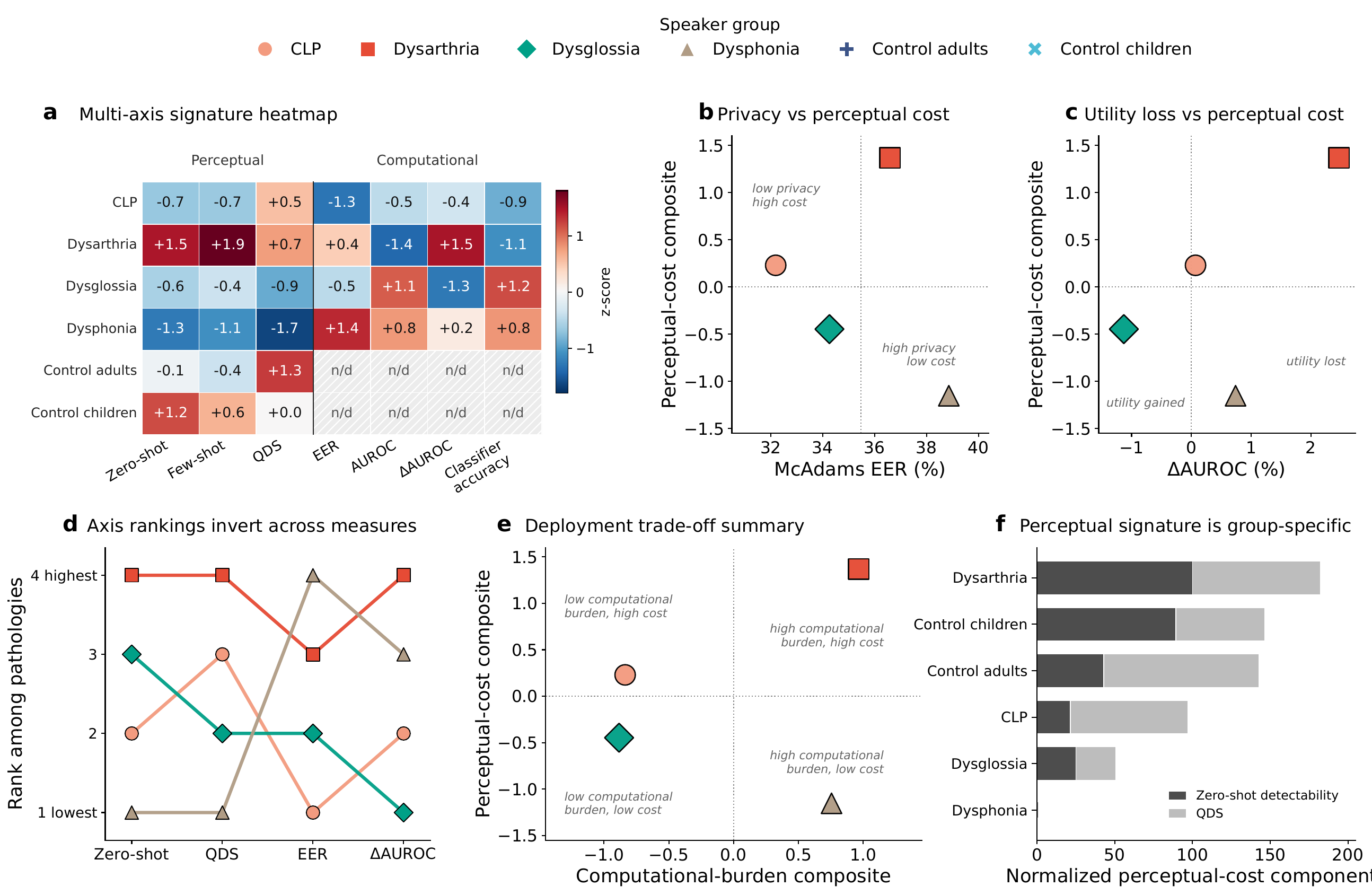}
\caption{Disorder-specific perceptual signatures emerge across all measured axes.
Perceptual measures are computed here for all six speaker groups; computational measures (EER, AUROC, $\Delta\text{AUROC}$, classifier accuracy) are defined only for the four pathology groups, because the downstream classifier treats controls solely as the negative class.
\textbf{a} Heatmap of the six groups across seven measures: zero-shot accuracy, few-shot accuracy, QDS, EER, post-anonymization AUROC, $\Delta\text{AUROC}$ (original minus anonymized), and post-anonymization classifier accuracy. Cells are $z$-normalized within column; computational cells for the two control groups are undefined and shown hatched. n/d, not defined.
\textbf{b},\textbf{c} Per-pathology perceptual-cost composite (mean of the within-group $z$-normalized zero-shot detectability and QDS) against EER (\textbf{b}) and against $\Delta\text{AUROC}$ (\textbf{c}). Thin lines mark the axis means (\textbf{b}) and $\Delta\text{AUROC}=0$ with the composite mean (\textbf{c}); each dot is one pathology group.
\textbf{d} Rank-inversion plot: each pathology's rank on zero-shot detectability, QDS, EER, and $\Delta\text{AUROC}$. Crossing lines indicate that a group's rank on one axis does not carry over to the others.
\textbf{e} Parallel-coordinates profile of the four pathology groups across five axes, each min-max scaled across the four groups.
\textbf{f} Per-group perceptual signature: horizontal bars decomposing each group's perceptual cost into its zero-shot detectability and QDS components, both min-max normalized across the six groups and stacked, sorted by total.}
\label{fig:signatures}
\end{figure}

Detectability, perceived quality loss, computational privacy gain, and downstream utility loss arrange the six speaker groups into distinct multi-metric profiles. No single axis collapses these dimensions: each pathology occupies its own region of the joint space, and the four pathologies for which paired computational measurements exist (CLP, Dysarthria, Dysglossia, Dysphonia) span the full range of high-low combinations along each axis (Fig.~\ref{fig:signatures}a). The rank a pathology holds on one axis does not predict its rank on the others, and several rankings invert outright between the perceptual and the computational measures (Fig.~\ref{fig:signatures}d).

The two extremes sit on opposite corners. Dysarthria is the most perceptually detectable group (zero-shot 96, QDS 34) and suffers the largest downstream utility loss of the four pathologies ($\Delta\text{AUROC} = 2.47$\%): the McAdams shift is conspicuous to listeners, lowers perceived quality substantially, and visibly degrades the classifier. Dysphonia occupies the opposite corner, with the lowest perceptual detectability of the cohort (86 zero-shot) and the smallest QDS (23), yet the highest McAdams EER (39\%) and only a modest AUROC drop (0.74\%). A parsimonious explanation combines a perceptual floor effect intrinsic to dysphonic voices, whose lower naturalness baseline leaves less headroom to degrade, with the fact that the laryngeal substrate of Dysphonia is acoustically distinct from the supraglottal formants the McAdams algorithm manipulates, so the disorder signature and the anonymization signature occupy largely non-overlapping spectral regions.

The two intermediate pathologies fill the space with their own configurations (Fig.~\ref{fig:signatures}b,c). CLP combines moderate detectability (88 zero-shot) and a high QDS (33) with the lowest EER of the four groups (32\%) and a near-zero AUROC drop (0.07): the algorithm provides the weakest privacy protection here while still imposing a sizeable perceived cost, plausibly because the pediatric CLP cohort has a different formant geometry from the adult voices that drive the ASV baseline. Dysglossia is the quietest signature on the computational side: its $\Delta\text{AUROC}$ is slightly negative ($-1.13$\%), so the downstream classifier marginally improves after anonymization, while detectability (89) and QDS (26) are mid-to-low; the McAdams shift appears to remove a speaker-identity confound that had been noise for the pathology classifier, without imposing a large perceptual change. The two control groups close the picture from a third angle: child controls reach the second-highest detectability of all six groups (95 zero-shot) and adult controls carry the largest QDS of any group (36). The transformation is therefore neither cosmetically nor perceptually transparent on healthy voices, and the high detectability of child controls in particular rules out any claim of perceptual transparency framed at the cohort level.

These contrasts argue against any universal characterization of the automatic anonymization as either high- or low-impact. The same algorithm, applied to acoustically distinct disorders, produces qualitatively distinct combinations of privacy gain, utility cost, and perceptual visibility (Fig.~\ref{fig:signatures}e,f). A deployment evaluation that ignores any one of these axes risks endorsing the algorithm for disorders on which it is in fact conspicuous to listeners, such as Dysarthria and child controls, or rejecting it on disorders where it is perceptually quiet, such as Dysphonia and Dysglossia. The pathology-specific perspective is the appropriate granularity at which to license clinical use of speech anonymization in pathological populations, and the multi-axis signature shown here can serve as a template for evaluating future anonymization methods on the same cohort.

\section*{Discussion}

This study presents a human-centered evaluation of automatically anonymized pathological speech, combining four perceptual tasks across 180 speakers sampled from a German clinical speech corpus \cite{maier2009peaks,arasteh2023effect,arasteh2024addressing}. Using the McAdams coefficient-based transformation \cite{mcadams1984spectral,patino2021speaker}, which has previously been characterized only on automatic privacy and utility metrics on the same pool \cite{arasteh2024addressing}, we measured zero-shot Turing-style discrimination, few-shot discrimination after brief familiarization, 5-point quality rating, and 4-point blinded clinical severity rating across a listener panel structured by native language and domain expertise. The discrimination tasks asked whether listeners can tell that a recording has been anonymized, not whether they can identify the speaker; the severity task asked whether a clinician's grade survives the transformation. Anchored on these four axes, the findings expose a structure that no single computational summary captures: the McAdams transformation is perceptually conspicuous, lowers perceived quality substantially, and yet preserves the clinical diagnostic signal almost intact, with effects that differ systematically by disorder and by listener attribute.

Listeners detected the presence of anonymization at 91\% accuracy on first exposure and 93\% after a brief familiarization phase, with the cue accessible without training. This rate is consistent with reports that even straightforward speech transformations leave residual perceptual signatures audible to attentive listeners \cite{rusz2021speech,fagherazzi2021voice}, and it contradicts an implicit assumption in the speaker-anonymization literature \cite{tomashenko2024voiceprivacy,srivastava2022privacy,fang2019speaker,patino2021speaker} that an automatically anonymized recording is, for practical purposes, perceptually transparent. Detectability varied significantly across pathologies in the zero-shot task: Dysarthria and child controls were the most easily identified as anonymized, Dysphonia the hardest. The variation tracks the acoustic substrate that the McAdams algorithm manipulates. Dysarthric speech carries broad-spectrum articulatory and prosodic deviations \cite{darley1969differential,kim2010dysarthria} that interact visibly with a formant-shifting pipeline, while dysphonic voices are dominated by laryngeal-source irregularities \cite{hirano1981clinical} whose perceptual identity is determined more by source noise and periodicity than by the supraglottal formant geometry the algorithm rotates. CLP and Dysglossia, which manipulate resonance and articulation rather than source, sit between the two extremes. The pattern argues against treating "anonymized speech" as a single perceptual object and in favor of evaluating each disorder against the specific spectral region the anonymization touches.

Few-shot familiarization narrowed but did not erase the cohort-level detection signal: overall accuracy rose by about two percentage points (paired $p = 0.015$), the across-group difference no longer reached significance, and at the individual cell level only 25 of 60 listener-by-group cells improved with familiarization while 26 were unchanged and 9 declined. This pattern is consistent with perceptual recalibration on a familiar transformation rather than the discovery of a new cue \cite{adank2010comprehension,bradlow2008perceptual}; the anonymization signature is largely accessible at first exposure, and familiarization smooths inter-pathology variance rather than creating a qualitatively different listening strategy. In deployment terms, a perceptual evaluation of a single anonymization pipeline cannot rely on naive listeners alone, since the result obtained at first exposure is not the result a clinician will report after working with the same pipeline for a week.

Two listener attributes shaped perception with opposite profiles. Native German speakers detected anonymization more accurately than non-native speakers (Cohen's $d = 0.76$ zero-shot, $d = 0.60$ few-shot), while domain expertise did not (both $|d| \leq 0.11$). Conversely, on perceived quality, expertise produced a large effect ($d = -0.78$, experts reported less degradation) while native language produced only a small, non-significant difference ($d = 0.42$, $p = 0.108$). This double dissociation places the two attributes on different functional dimensions: detection of the anonymization signature is anchored in low-level phonemic and prosodic familiarity with the native language \cite{bradlow1997intelligibility,cooke2008foreign,best1995direct}, while the magnitude of perceived degradation is calibrated by clinical or signal-processing training \cite{kreiman1993perceptual,oates2009auditory}. The practical consequence is direct: a perceptual evaluation that recruits only one type of listener will systematically misjudge one of the two outcomes, overstating quality degradation if it uses only non-experts and understating detectability if it uses only non-native listeners. Reporting both subgroups is the minimum standard for a perceptual evaluation of a clinical anonymization pipeline.

Speaker-side demographics, in contrast, produced no detectable bias. Female and male speakers were discriminated and rated for quality at indistinguishable rates across the cohort, and the result held within every individual pathology group (no within-group sex contrast reached significance across 18 tests, the smallest $p$-value being 0.097). This is reassuring even for Dysphonia, whose strongly male-skewed composition (25 male, 5 female) could in principle have masked or exaggerated apparent disparities. Speaker age was uncorrelated with detectability and showed only a shallow negative association with quality degradation that strengthened slightly in adults only ($\rho = -0.302$, $p = 0.001$), plausibly a within-cohort echo of the floor effect seen across pathologies: voices that already sit lower on the naturalness axis have less headroom to lose under transformation. Together, the two demographic findings indicate that the McAdams transformation does not advantage one sex or age stratum over another, a fairness property \cite{morales2020sensitivenets} that is required, not optional, in clinical populations whose demographic composition is rarely uniform.

Beyond detectability, the transformation produced a substantial drop in perceived quality, with 175 of 180 speakers (97\%) showing a positive mean quality degradation score and the cohort-mean drop reaching 30 percentage points on the 0--100 scale (paired $p < 0.001$). The magnitude was pathology-dependent. The largest drops occurred in adult controls, Dysarthria, and CLP, and the smallest in Dysphonia and Dysglossia, with seven of fifteen pairwise contrasts surviving FDR correction along the same gradient. The asymmetry produces a marked reorganization of the perceived-quality hierarchy after anonymization: Dysphonia, which had been the second-lowest-rated group in the originals, ends with the highest mean rating, and CLP, which had been mid-pack, ends with the lowest. The standard deviation across group means falls from 4.31 to 3.35 (ratio 0.78), indicating partial convergence of inter-group quality once the anonymization signature is overlaid. A parsimonious interpretation is a perceptual floor effect: pathologies whose original recordings already sit lower on the naturalness axis have less headroom to lose, and the McAdams shift is rated more leniently relative to a degraded baseline. The clinical implication is that group-level "quality after anonymization" loses some of its diagnostic resolution; researchers studying how anonymized speech compares across disorders should be aware that the transformation itself partly compresses the very contrasts they may wish to measure.

The clinically central finding, however, is that pathological severity was preserved at the level required for diagnostic interpretation. Across all 180 speakers, no recording shifted by more than one grade between original and anonymized on the 0--3 ordinal scale; 80\% of speakers received identical grades; and among 71 pathology speakers originally rated as having any pathology (severity $\geq 1$), 69 (97\%) remained at grade 1 or higher after anonymization. Per-group exact-grade agreement was 87\% for Dysarthria, 87\% for Dysglossia, and 90\% for Dysphonia, with quadratic-weighted Cohen's $\kappa$ \cite{cohen1968weighted} of $0.87$, $0.93$, and $0.94$ for these three groups, all in the conventional "almost perfect" range \cite{landis1977measurement}. The two outlier patterns were both upward and both clinically benign. CLP was the only pathology group with lower agreement ($\kappa_w = 0.57$), driven mostly by mild-grade upgrades of originally-grade-0 recordings, and the two control groups showed an analogous mild over-call (23\% of healthy recordings newly flagged as grade 1 after anonymization). The pattern is consistent with the residual McAdams noise being read as mild hoarseness on a clinical scale that does not distinguish source noise from formant artifact \cite{hirano1981clinical}, an interpretive caveat the rating clinician explicitly raised. Crucially, the clinically dangerous direction, an anonymized pathological recording being read as healthy, occurred in only 2 of 120 pathology speakers, both mild-grade CLP; severity escalation beyond mild within the pathology cohort did not exceed 3 of 39 originally-mild speakers. The transformation is therefore perceptually conspicuous and lowers perceived speech quality, but it leaves the diagnostic pathology signal substantively intact at the pathology level and within each clinical group.

The most surprising finding is that the perceptual and computational evaluations of the same transformation are decoupled. Across the four pathology groups with paired computational measurements, the McAdams EER did not track zero-shot detectability ($r = +0.028$), few-shot detectability ($r = +0.092$), or quality degradation ($r = -0.546$). The dissociation is sharpest at the two extremes: Dysphonia carries the highest McAdams EER (the strongest computational anonymization) yet the lowest perceptual detectability and the smallest QDS in the cohort, while CLP shows the inverse arrangement. An evaluator relying on EER alone would rank Dysphonia the cleanest application of the algorithm and CLP the weakest, whereas the perceptual evidence reverses that ordering on QDS and equalizes the two on detectability. Utility change ($\Delta\text{AUROC}$) tracked perception slightly better in direction (positive correlations with both detectability and QDS) but never significantly so at the four available group means, and the per-pathology correlation matrix between three perceptual and four computational measures contains no significant cell. The current standard in the speaker-anonymization literature \cite{tomashenko2024voiceprivacy,patino2021speaker,fang2019speaker,srivastava2022privacy} of validating a pipeline purely on EER and AUROC, taken alone, would judge all four pathologies broadly successful applications of the McAdams method. The perceptual evidence shows instead that the same transformation is highly conspicuous in Dysarthria, moderately so in CLP, and comparatively unobtrusive in Dysphonia and Dysglossia. Computational and perceptual evaluations therefore index different facets of the transformation: speaker-identity confusability on one side, naturalness change and diagnostic preservation on the other. A clinically responsible release of anonymized speech should report both.

Taken together, detectability, perceived quality loss, computational privacy gain, and utility loss arrange the six speaker groups into multi-metric profiles that do not collapse onto a single axis. The two extremes occupy opposite corners. Dysarthria is the most perceptually detectable, suffers the largest quality drop, and incurs the largest downstream utility loss; the algorithm is conspicuous on every axis. Dysphonia is the inverse: the lowest perceptual detectability and the smallest QDS in the cohort, yet the highest McAdams EER and only a modest AUROC drop. Between them, CLP combines moderate detectability and a high QDS with the lowest EER and a near-zero AUROC drop, the unfavorable corner of weak privacy gain at sizeable perceived cost, plausibly because the pediatric CLP cohort has a different formant geometry from the adult voices that drive the automatic speaker-verification baseline; Dysglossia is the quietest signature on the computational side ($\Delta\text{AUROC} = -1.13$\%, the downstream classifier marginally improves after anonymization) at mid-low perceptual cost, an arrangement in which the McAdams shift removes a speaker-identity confound that had been noise for the pathology classifier without imposing a large perceptual change. The two control groups close the picture from a third angle: child controls reach the second-highest detectability of all six groups and adult controls carry the largest QDS of any group, ruling out any claim of perceptual transparency framed at the cohort level. 

It is important to acknowledge that anonymization necessarily alters acoustic structure, and use cases that critically depend on fine-grained acoustic measurements, such as treatment-outcome evaluation, longitudinal phonatory analysis, or quantitative voice assessment with cepstral, spectral, and prosodic markers \cite{maryn2009acoustic,maryn2010acoustic,laver1981acoustic,patel2018recommended,fant1970acoustic,kummer2013speech}, may be incompatible with anonymized recordings. Perceptual judgments alone, as gathered here, are not a substitute for those measures. The contribution of this study is complementary: it characterizes the perceptual signal a clinician or researcher will encounter when consuming an anonymized recording, and it ties that signal to the privacy and utility metrics that already drive deployment decisions. Anonymized recordings can still support method development, algorithm benchmarking, perceptual studies, educational use, and exploratory analyses where privacy constraints would otherwise preclude data sharing, and the present findings allow a researcher to estimate, for each disorder, what is gained and what is lost by accepting an anonymized version of the data.

This study has several limitations. First, the listener panel was small ($n = 10$), and the non-native subgroup spans a wide range of German proficiencies and language families. We deliberately recruited a diverse interdisciplinary panel to reflect realistic research-use scenarios at the cost of statistical power on each subgroup-by-task combination; the cell-level analyses used throughout mitigate this by recovering 30 to 60 measurements per subgroup contrast. Second, the severity ratings rest on a single phoniatrician, which is standard for blinded clinical scoring on this scale but means that the reported severity-preservation result has not been replicated across clinicians. Expansion to a multi-rater clinical panel with inter-rater agreement on both conditions is a natural next step and would be the appropriate study to license severity-preservation claims at scale. Third, all speakers were German, and the McAdams transformation interacts with formant geometry that is partly language-specific \cite{patino2021speaker}; cross-linguistic replication on pathological corpora in other languages is required before the disorder-specific signatures reported here are generalized. Fourth, the perceptual-computational alignment uses four pathology means, which are the inferential limit of any contrast at this granularity; the correlations reported there are descriptive of the trend rather than inferential of the magnitude, and we have framed them as such throughout. Fifth, we evaluated a single anonymization pipeline (McAdams) at a single operating point. The systematic pathology-by-axis structure documented here predicts that other anonymization families, such as neural voice-conversion or signal-processing alternatives, will produce qualitatively different perceptual signatures, and the same four-task protocol can be reused to characterize them. Sixth, one listener used hearing aids during the evaluation, which we expect to have negligible effect given the structured randomization but cannot rule out without replication \cite{lunner2009cognition}. Finally, our group definitions emphasized dominant acoustic manifestations rather than mutually exclusive etiologies; diagnostic overlap between dysarthric, dysphonic, and dysglossic speech is a clinical reality \cite{darley1969differential,kummer2013speech,hirano1981clinical} that the cohort partially absorbs and that a clinical reader should keep in mind when interpreting the per-group differences.

The broader picture from this work is that automatic anonymization of clinical speech makes a trade-off that cannot be assessed on privacy and utility metrics alone. The McAdams coefficient method is perceptually salient to listeners, lowers perceived quality substantially, and yet leaves the diagnostic pathology signal almost intact; its disorder-specific footprint runs in different directions on computational and perceptual axes; and its effects on individual listeners depend systematically on native language and on clinical training in different ways. None of these facts are visible in EER and AUROC alone. A clinically responsible evaluation of speech anonymization in pathological populations should report all four axes, on disorder-stratified data, with at least one trained clinician scoring severity under blinding. The protocol used here is one concrete way to do that, and the cohort and results are made available so that future anonymization methods can be evaluated against the same reference points. Privacy and clinical interpretability are not on the same axis; only an evaluation that measures both can certify that one was not bought at the price of the other.



\section*{Materials and Methods}

\subsection*{Ethics statement}

The study and the methods were performed in accordance with relevant guidelines and regulations and approved by the University Hospital Erlangen’s institutional review board with application number 3473. Informed consent was obtained from all adult participants as well as from parents or legal guardians of the children. All audio data used in this study were de-identified prior to listener access. The evaluation protocol adhered to ethical guidelines for perceptual studies involving anonymized speech and received internal approval for data handling and experimental procedures. Participation by expert listeners was voluntary and non-incentivized, and all participants provided informed agreement to take part in the listening tasks.


\subsection*{Clinical speech corpus and study cohort}

The study used an internal German clinical speech corpus collected within the PEAKS framework, the Program for Evaluation and Analysis of all Kinds of Speech disorders \cite{maier2009peaks}. The corpus contains more than 200 hours of recordings from over 2700 native German speakers and spans an age range from 3 to 95 years \cite{maier2009peaks,arasteh2024addressing,arasteh2023effect,tayebi2025differential}. Recordings were acquired between 2006 and 2019 during routine outpatient examinations at the University Hospital Erlangen and at more than 20 additional sites across Germany. The corpus includes healthy speakers and speakers with clinically documented speech or voice disorders, including Cleft lip and palate (CLP) \cite{millard2001different,harding1996characteristics,maier2006fully}, Dysarthria \cite{hirose1986pathophysiology}, Dysglossia \cite{schroter2005rehabilitation}, and Dysphonia \cite{sama2001clinical}. The original recordings were stored as 16-bit audio sampled at 16 kHz. As described previously \cite{arasteh2023effect}, recordings were collected using a small set of headset microphones that differed by speaker group: the dnt Call 4U Comfort microphone \cite{maier2006fully} was used for Dysglossia recordings, a Plantronics headset \cite{console46plantronics} was used for Dysarthria, CLP, adult controls, and child controls, and a Logitech headset \cite{logitechk} was used for Dysphonia recordings.

From this corpus, we derived the analytic cohort used in the present perceptual study. The cohort was selected by stratified sampling from eligible recordings and contained 180 speakers in total, with six groups of 30 speakers each: CLP, Dysarthria, Dysglossia, Dysphonia, adult controls, and child controls. The same 180 speakers were used for the Turing-style discrimination experiments, the subjective quality-rating experiment, and the expert clinical severity-preservation experiment. For each speaker, one original utterance and one anonymized counterpart were retained, yielding 180 original-anonymized pairs and 360 audio files in total. Across the selected cohort, mean age was 35 $\pm$ 24 years, with a range from 6 to 78 years, and the cohort included 93 female and 87 male speakers. The demographic composition of the speaker cohort and the listener panel is summarized in Table~\ref{tab:cohort}.

Speakers were eligible for the analytic cohort if they were native German speakers, had a clinically documented group assignment, had standardized speech material suitable for the listening protocol, and had recordings of sufficient quality for perceptual evaluation. Recordings were excluded before sampling if they had ambiguous or mixed diagnostic labels, non-standardized speech material, insufficient audio quality, or metadata that made them unsuitable for the blinded perceptual experiments. Diagnostic groups were assigned according to the dominant clinical documentation available in the PEAKS corpus rather than by re-diagnosis for the present study. The groups should therefore be interpreted as clinically documented dominant categories rather than mutually exclusive biological entities, since overlap between speech and voice symptoms can occur in real clinical populations.

The four pathology groups were retained because they represent different dominant mechanisms of speech or voice impairment. CLP is mainly associated with resonance and articulation changes related to cleft lip and palate, including hypernasality and compensatory articulation patterns \cite{harding1996characteristics,maier2006fully}. Dysarthria reflects motor speech impairment due to impaired neuromuscular control and can affect articulation, prosody, phonation, and speech rhythm \cite{hirose1986pathophysiology}. Dysglossia reflects articulatory impairment related to structural or peripheral abnormalities of the speech organs \cite{schroter2005rehabilitation}. Dysphonia reflects laryngeal voice impairment, primarily affecting phonation and perceived voice quality \cite{sama2001clinical}. The control groups were retained to separate pathology-related effects from the perceptual consequences of anonymization in speakers without documented speech or voice pathology. Child controls provide the age-appropriate comparison for the pediatric CLP group, whereas adult controls provide a non-pathological reference for the adult speech material. The adult control group was younger than the adult pathology groups in the selected cohort, as shown in Table~\ref{tab:cohort}; therefore, age was analyzed explicitly as a speaker-side covariate in the demographic analyses rather than assumed to be fully matched across adult groups.

No additional loudness equalization or acoustic normalization was applied before constructing the perceptual stimuli, so that the listening experiments preserved the acoustic conditions of the clinical recordings. For the perceptual experiments, listeners received only de-identified audio files without speaker names, clinical metadata, demographic metadata, recording-site information, or anonymization labels.

\subsection*{Speech tasks, recording conditions, and utterance selection}

The perceptual experiments used short excerpts derived from standardized speech material in the PEAKS corpus. Speakers in the Dysarthria, Dysglossia, Dysphonia, and adult-control groups produced the German version of the phonetically balanced passage Der Nordwind und die Sonne (The North Wind and the Sun), which contains 108 words and is commonly used in clinical speech assessment within this corpus \cite{maier2009peaks}. Speakers in the pediatric groups, namely CLP and child controls, completed the Psycholinguistische Analyse kindlicher Sprechstörungen (PLAKSS) picture-naming task, which was designed for the assessment of developmental speech disorders in German-speaking children \cite{fox2002plakss}. The use of different speech tasks followed the structure of the underlying clinical corpus and ensured that each age group was represented by task material appropriate to the original clinical assessment.
The recording conditions were inherited from the clinical corpus. We did not apply additional loudness normalization, noise reduction, or post hoc acoustic equalization before the perceptual experiments, because the aim was to evaluate anonymization under the same acoustic conditions as the available clinical recordings.

For the present study, one utterance of approximately 3 to 4 seconds was selected for each speaker. In the pediatric groups, where the PLAKSS task consists of shorter picture-naming responses and natural pauses between productions, recordings were segmented at pauses longer than 1 second before selecting the final excerpt \cite{maier2009peaks}. For the passage-reading groups, the selected excerpt was taken from the available standardized passage recording. The final selection produced one original utterance per speaker, after which each utterance was anonymized once using the McAdams coefficient method. This yielded a paired design in which every speaker contributed exactly one original file and one anonymized counterpart to all perceptual experiments.

The selection was constrained by the requirements of the listening protocol. Each excerpt had to contain usable speech, be long enough to support perceptual judgment, and be short enough to keep the full experiment feasible for repeated human evaluation. The same utterance pair was used across the zero-shot discrimination task, the few-shot discrimination task, the quality-rating task, and the expert clinical severity-preservation task, so that differences between experiments reflected the rating instructions rather than changes in stimulus material.


\subsection*{Automatic speech anonymization}

All selected utterances were anonymized with the McAdams coefficient method used in our previous work on the same clinical speech corpus \cite{arasteh2024addressing}. The present study does not introduce a new anonymization algorithm; instead, it evaluates the perceptual and clinical consequences of applying this fixed anonymization procedure to pathological speech. The full technical formulation of the method, its relation to signal-processing and neural anonymization families, and implementation details are provided in Supplementary Note~\ref{supnote:anonymization_method}.

Briefly, the McAdams method is a deterministic signal-processing approach for speaker anonymization \cite{arasteh2024addressing,mcadams1984spectral,patino2021speaker}. It applies linear predictive coding to each speech frame and modifies the angular frequencies of the complex poles of the vocal-tract filter. For a linear predictive coding pole written as $z_k = r_k e^{j\phi_k}$, the pole angle is transformed as $\phi_k' = \phi_k^{\alpha},$
where $\alpha$ is the McAdams coefficient. This operation shifts the spectral envelope and formant structure, which contribute to perceived speaker identity, while preserving the original residual excitation. The method therefore changes the apparent voice characteristics without mapping the speaker to a target or pseudo-speaker identity. For each of the 180 selected utterances, one anonymized counterpart was generated and used unchanged in the blinded perceptual experiments.


\subsection*{Blinded stimulus preparation}

All perceptual experiments used the same 180 original-anonymized utterance pairs. The stimuli were arranged into six blinded sets, each containing 30 speakers from one clinical or control group. The group structure was preserved for analysis, but the task-facing set names did not reveal the diagnostic category. For each speaker, the original utterance and its anonymized counterpart were presented as two numbered files. The assignment of original and anonymized recordings to file position was randomized and concealed from the listeners. The private key linking speaker group, speaker identity, file position, and anonymization status was retained only for statistical analysis.

In the discrimination experiments, each trial contained the two recordings from the same speaker, and listeners indicated which recording they perceived as the original. In the quality-rating experiment, the same recordings were rated for naturalness and audio quality without revealing anonymization status. The zero-shot discrimination task was completed before any repeated-listening task, preventing later familiarization from influencing first-exposure judgments. The few-shot discrimination and quality-rating tasks were then completed under the same blinded file assignment, allowing repeated listening while preserving the original-anonymized mapping for paired analysis.

The expert clinical severity-preservation experiment used an analogous blinded organization derived from the same 180 pairs. The expert rater evaluated the two files from each speaker independently and assigned a perceived pathological severity score to each file. The instructions explicitly discouraged comparison-based judgments of relative severity and discouraged attempts to identify which file was original or anonymized. This design allowed severity preservation to be assessed as a blinded clinical-perceptual property of each recording rather than as a preference or detection task.


\subsection*{Listener panel}

Ten listeners participated in the main perceptual evaluation. The panel was designed to vary along two listener-side attributes: native-language background and domain expertise. Five listeners were native German speakers and five were non-native German speakers. The non-native listeners reported German proficiency levels from A1 to C1 according to the Common European Framework of Reference for Languages \cite{little2020common}. The native German listeners reported native-level fluency. The listener composition is summarized in Table~\ref{tab:cohort}.
Domain expertise was defined as at least 8 years of specialized experience in signal-level speech processing or in clinical voice and speech assessment. Under this definition, four listeners were classified as experts: three technical experts in speech processing and one clinical expert in phoniatrics. The clinical expert, L8, was the only phoniatrician in the panel and was therefore also selected for the additional clinical severity-preservation experiment. The remaining six listeners were classified as non-experts for the purposes of the listener-attribute analyses. L7 was a clinician but specialized in radiology rather than voice, speech, or phoniatrics, and was therefore treated as a non-expert in the present study.

All ten listeners completed the zero-shot discrimination, few-shot discrimination, and quality-rating tasks. They were blinded to speaker identity, diagnostic group, speaker sex, speaker age, recording site, microphone type, and anonymization status. The listeners were instructed to base their judgments only on the audio files and to complete the zero-shot task before any repeated-listening task, so that first-exposure discrimination could be analyzed independently from familiarization effects.


\subsection*{Perceptual discrimination experiments}

The perceptual detectability of anonymization was evaluated with a two-alternative forced-choice, Turing-style discrimination paradigm \cite{turing1950computing}. Each trial contained the two recordings from the same speaker, one original and one anonymized, and the listener was asked to identify which recording was the original. The task was not designed to measure whether listeners could recognize the speaker's identity after anonymization. Instead, it measured whether the acoustic transformation introduced by anonymization was perceptually detectable when the original and anonymized recordings were presented as a matched pair.

\paragraph{Zero-shot discrimination.}

In the zero-shot condition, listeners evaluated all 180 speaker pairs without prior familiarization. For each speaker, the listener heard the two recordings only once and then selected the file perceived as the original recording. The zero-shot task was completed before any repeated-listening or quality-rating task. This ordering was used to capture first-exposure sensitivity to the anonymization transformation and to prevent later familiarity with the stimulus set from influencing the zero-shot judgments. Each listener contributed 180 binary decisions in this condition, corresponding to 30 decisions per speaker group and 1800 zero-shot decisions across the full listener panel.

\paragraph{Few-shot discrimination.}

After completing the zero-shot task, listeners performed the few-shot discrimination task on the same 180 speaker pairs under the same blinded assignment of original and anonymized file positions. In this condition, listeners could listen to the two recordings from each speaker as many times as needed before selecting the file perceived as the original. The few-shot task was included to test whether repeated exposure increased sensitivity to the anonymization transformation or altered the pathology-dependent pattern observed under first exposure. Each listener again contributed 180 binary decisions, yielding 1800 few-shot decisions across the full panel.

\paragraph{Discrimination accuracy.}

Discrimination accuracy was defined as the proportion of trials in which the listener correctly identified the original recording within an original-anonymized pair. For a listener or analysis stratum with $N$ trials, accuracy was computed as

\begin{equation}
\mathrm{Accuracy} =
\frac{1}{N}\sum_{i=1}^{N} \mathbb{I}(\hat{y}_i = y_i) \times 100,
\label{eq:discrimination_accuracy}
\end{equation}
where $y_i$ denotes the true file position of the original recording in trial $i$, $\hat{y}_i$ denotes the listener's response, and $\mathbb{I}(\cdot)$ is the indicator function. Because each trial had two possible responses, chance performance was 50. Accuracy was computed per listener, per speaker group, per listening condition, and for listener-attribute and speaker-demographic strata used in the statistical analyses.


\subsection*{Subjective speech-quality rating experiment}

The subjective quality-rating experiment quantified how anonymization changed the perceived naturalness and audio quality of the recordings. This task was distinct from the discrimination experiment. In the discrimination task, listeners judged which file was original within a matched pair; in the quality task, they assigned an absolute quality score to each recording under blinded conditions. The quality ratings were therefore used to measure perceived signal degradation rather than anonymization detectability or speaker identifiability.

\paragraph{Quality-rating task.}

After completing the zero-shot discrimination task, listeners rated the original and anonymized recordings from all 180 speakers. Each listener rated 360 audio files in total, corresponding to one original and one anonymized file for each speaker. Ratings were given on a five-point ordinal scale, where 1 indicated the lowest perceived naturalness and audio quality and 5 indicated the highest perceived naturalness and audio quality. Listeners were reminded that the recordings came from speakers with speech or voice disorders and that reduced intelligibility or atypical speech characteristics could be present in the original clinical recordings. The task was therefore framed as a judgment of perceived naturalness and audio quality, not as a diagnostic judgment, intelligibility test, or severity rating.

\paragraph{Normalized quality score.}

For reporting and visualization, the five-point quality scores were linearly mapped to a 0--100 scale, with 1 mapped to 0 and 5 mapped to 100. For a set of $N$ ratings with raw score $s_i \in \{1,2,3,4,5\}$, the normalized quality score was computed as

\begin{equation}
Q =
\frac{1}{N}\sum_{i=1}^{N}
\frac{s_i - 1}{4}
\times 100.
\label{eq:normalized_quality}
\end{equation}
This transformation was used only to make the magnitude of quality differences easier to interpret and to keep the reporting scale consistent with the discrimination results. The underlying listener response remained an ordinal five-point rating. Normalized quality scores were computed separately for original and anonymized recordings, for each listener, speaker group, and analysis stratum.

\paragraph{Quality degradation score.}

The quality degradation score (QDS) measured the perceived quality loss introduced by anonymization. For each listener and speaker, QDS was defined as the normalized quality score of the original recording minus the normalized quality score of the anonymized counterpart,

\begin{equation}
\mathrm{QDS}_{j,k} =
Q^{\mathrm{orig}}_{j,k} - Q^{\mathrm{anon}}_{j,k},
\label{eq:qds}
\end{equation}
where $j$ indexes the listener and $k$ indexes the speaker. Positive values indicate that the anonymized recording was rated lower than the original recording, a value of 0 indicates no perceived quality change, and negative values indicate that the anonymized recording was rated higher than the original. QDS was computed at the listener-speaker level before aggregation, preserving the paired structure of the original and anonymized recordings.


\subsection*{Expert clinical severity preservation experiment}

The clinical severity-preservation experiment was added to test whether anonymized recordings retained the perceptual evidence of pathological impairment. This experiment directly differed from the discrimination and quality-rating tasks. The expert was not asked to detect anonymization, identify the original file, or judge whether the anonymized output was acceptable. Instead, each file was rated independently for the perceived severity of speech or voice impairment.

\paragraph{Blinded severity-rating dataset.}

The severity experiment used the same 180 original-anonymized utterance pairs as the main perceptual experiments. The stimuli were arranged into six blinded sets with 30 speakers per set and two files per speaker. One file was the original recording and the other was its anonymized counterpart, but the file assignment was concealed from the expert rater. The blinded set names did not reveal the clinical or control category. The private key linking set identity, speaker identity, file position, diagnostic group, and anonymization status was retained only for analysis after completion of the ratings.

\paragraph{Clinical severity-rating task.}

The task was completed by the clinical phoniatrician in the listener panel (M.S. with 35 years of experience). The listener rated each audio file independently using a four-level ordinal scale for perceived pathological impairment, where 0 indicated no pathological impairment, 1 mild impairment, 2 moderate impairment, and 3 severe impairment. The instructions explicitly stated that each file should be judged only on the basis of what was heard in that file. The expert was instructed not to infer which file was original or anonymized, not to compare the two files from the same speaker to decide relative severity, and not to base the score on perceived anonymization quality or naturalness. The target judgment was only the severity of the perceived speech or voice impairment.

\paragraph{Severity preservation outcomes.}

Let $S^{\mathrm{orig}}_k$ and $S^{\mathrm{anon}}_k$ denote the expert severity scores for the original and anonymized recordings of speaker $k$, with $S \in \{0,1,2,3\}$. The primary severity-preservation outcome was the paired score difference,

\begin{equation}
\Delta S_k = S^{\mathrm{anon}}_k - S^{\mathrm{orig}}_k ,
\label{eq:severity_difference}
\end{equation}
where $\Delta S_k = 0$ indicates unchanged perceived severity after anonymization, negative values indicate a lower severity rating after anonymization, and positive values indicate a higher severity rating after anonymization. We also computed the absolute severity shift,

\begin{equation}
|\Delta S_k| = \left|S^{\mathrm{anon}}_k - S^{\mathrm{orig}}_k\right|,
\label{eq:absolute_severity_shift}
\end{equation}
to quantify the magnitude of severity change irrespective of direction. In addition, each file was converted to a binary pathological-status label using $S>0$ as pathological and $S=0$ as non-pathological. This allowed us to evaluate whether anonymized recordings retained the expert's perception of pathological vs non-pathological speech, separately from the finer ordinal severity score. Severity preservation was summarized for the full cohort and by speaker group using the proportion of speakers with unchanged severity score, the proportion with a one-level shift, the proportion with a shift of two or more levels, and the proportion with unchanged binary pathological status.


\subsection*{Computational reference metrics}

Computational privacy and utility metrics were used as reference measurements against which the human perceptual outcomes were compared. These metrics were taken from our previous computational study on the same clinical speech corpus and the same McAdams anonymization pipeline \cite{arasteh2024addressing}. They were not used to train, calibrate, or guide the listener experiments. Because the present paper focuses on human perception, the computational metrics served only as group-level reference axes for testing whether automatic anonymization performance aligns with perceptual detectability, perceived quality degradation, and clinical severity preservation.

\paragraph{ASV privacy metric.}

Privacy was quantified using the equal error rate (EER) \cite{hansen2015speaker} of an automatic speaker verification (ASV) system \cite{kinnunen2010overview}. EER is the operating point at which the false acceptance rate equals the false rejection rate; higher EER after anonymization indicates that the system is less able to verify speaker identity and therefore indicates stronger privacy protection \cite{7472729}. The ASV system used in the reference analysis consisted of three long short-term memory \cite{schmidhuber1997long} layers with 768 hidden units each, followed by a linear projection layer to obtain fixed-dimensional speaker embeddings \cite{sak2014long,arasteh2023effect,arasteh2024addressing}. The model was pretrained on LibriSpeech \cite{panayotov2015librispeech} using the generalized end-to-end ASV loss \cite{wan2018generalized,arasteh2022empirical} and optimized with Adam \cite{kingma2015adam}.

Input features for ASV were 40-dimensional log-Mel spectrograms extracted after voice-activity preprocessing \cite{ramirez2007voice}. Low-energy frames below 30 dB were discarded, silence was removed using a 30 ms window with a maximum allowable silence of 6 ms, and the short-time Fourier transform used a 25 ms window, a 10 ms hop, and a 512-point fast Fourier transform, following the reference implementation \cite{arasteh2023effect,arasteh2024addressing,wan2018generalized,arasteh2022empirical,prabhavalkar2015automatic}. Speaker similarity between an enrollment utterance and a verification utterance was computed by cosine similarity between their embeddings,

\begin{equation}
\mathrm{sim}(\mathbf{e}_{\mathrm{enroll}},\mathbf{e}_{\mathrm{verify}}) =
\frac{
\mathbf{e}_{\mathrm{enroll}}^\top \mathbf{e}_{\mathrm{verify}}
}{
\left\lVert \mathbf{e}_{\mathrm{enroll}} \right\rVert_2
\left\lVert \mathbf{e}_{\mathrm{verify}} \right\rVert_2
},
\label{eq:cosine_similarity}
\end{equation}
where $\mathbf{e}_{\mathrm{enroll}}$ and $\mathbf{e}_{\mathrm{verify}}$ denote the speaker embeddings of the enrollment and verification utterances. EER was obtained by sweeping the decision threshold over similarity scores and identifying the threshold at which false acceptance and false rejection rates were equal. In the present analysis, McAdams EER values were available for CLP, Dysarthria, Dysglossia, and Dysphonia and were compared with the corresponding group-level perceptual measures.

\paragraph{Pathology-classification utility metric.}

Utility preservation was quantified using the area under the receiver operating characteristic curve (AUROC) of pathology-specific binary classifiers from the same prior computational study \cite{arasteh2024addressing}. For each pathology group, a separate classifier was trained to distinguish pathological speech from healthy controls. These classifiers were trained and evaluated on the larger clinical corpus rather than on the 180-speaker perceptual subset, because the perceptual subset was designed for balanced human listening experiments and was too small to support robust classifier training.

The reference classifiers used log-Mel spectrograms as input rather than handcrafted acoustic features \cite{chlasta2019automated,muzammel2020audvowelconsnet,arasteh2024addressing}. Each input was represented by an 80-dimensional log-Mel spectrogram computed with a 1024-point fast Fourier transform. A forward-backward filter was applied when needed to suppress background drift \cite{gustafsson2002determining}. Spectrograms were reshaped into a three-channel format and passed to a ResNet-34 \cite{he2016deep} architecture pretrained on ImageNet \cite{deng2009imagenet}. The final layer was adapted for binary pathology classification. Training used binary weighted cross-entropy loss, the Adam optimizer, a learning rate of $5 \times 10^{-5}$, and a batch size of 8 \cite{kingma2015adam}. The input tensor size was $8 \times 3 \times 80 \times 180$, corresponding to batch size, channel number, Mel bins, and time frames.

For the reference analysis, speakers were split into speaker-disjoint training and test sets using a 70/30 split. To reduce class imbalance, the number of patient speakers for adult disorders was capped at twice the control-group size, while the number of controls in the CLP child subset was capped at 1.5 times the number of patients. The final training/test sizes were 168/73 for Dysarthria, 168/73 for Dysglossia, 110/49 for Dysphonia, and 887/381 for CLP. Each experiment was repeated over 50 randomized trials with paired evaluation of original and anonymized speech. AUROC was used as the primary computational utility metric. We additionally computed $\Delta\mathrm{AUROC}$ as original AUROC minus anonymized AUROC, so that positive values indicate reduced downstream classification utility after anonymization and negative values indicate higher classifier AUROC after anonymization. These utility metrics were available for the four pathology groups and were compared with the corresponding perceptual outcomes in the present study.


\subsection*{Statistical analysis}

All hypothesis tests were two-sided unless otherwise stated. The nominal significance threshold was $\alpha=0.05$. Benjamini-Hochberg FDR correction was applied to post-hoc pairwise group comparisons and reported as adjusted q-values. Descriptive results were summarized as mean $\pm$ standard deviation with 95\% confidence intervals (CIs). Listener-derived percentages were rounded to integers for reporting, and the percent sign was omitted in the Results except where needed for clarity. Statistical quantities, including p-values, correlation coefficients, and effect sizes, were reported to three decimal places. CIs for means were computed from the relevant analysis unit using the 95\% CI of the mean. For the main listener-level summaries, the analysis unit was the listener. For cell-level listener-attribute analyses, the analysis unit was the listener-by-group cell. For speaker-demographic analyses, the analysis unit was the speaker after averaging over listeners.

Perceptual discrimination was analyzed separately for the zero-shot and few-shot tasks. Group-level pathology effects were tested with repeated-measures one-way ANOVA \cite{OneWayANOVA,BasicandAdvancedStatisticalTests,anova_main,repeatedmeasures} on per-listener group accuracies, with speaker group as the within-listener factor. When the omnibus test indicated a group effect, pairwise group contrasts were performed with two-sided paired tests across listeners and corrected for multiple comparisons using the Benjamini-Hochberg false discovery rate (FDR) procedure \cite{benjamini1995controlling}. The overall change from zero-shot to few-shot discrimination was tested with a paired two-sided t-test on the 10 per-listener mean accuracies. The direction of the learning effect was additionally summarized at the listener-by-group level by counting cells in which few-shot accuracy increased, remained unchanged, or decreased relative to zero-shot accuracy.

Perceived speech quality was analyzed using the normalized 0--100 quality scale defined in Eq.~\ref{eq:normalized_quality}. The overall difference between original and anonymized recordings was tested with a paired two-sided t-test on per-listener mean quality scores. Group-level differences in original quality, anonymized quality, and quality degradation score were tested with repeated-measures one-way ANOVA on per-listener group means. Pairwise group contrasts for QDS were performed across listeners and corrected with FDR correction. The proportion of speakers with positive QDS was computed after averaging QDS over listeners for each speaker. Inter-listener agreement in perceived degradation was summarized by pairwise Spearman correlations \cite{spearman1961proof} between listeners using per-speaker QDS values.

Clinical severity preservation was analyzed after completion of the expert severity ratings. The primary ordinal outcome was the paired severity difference between anonymized and original recordings, defined in Eq.~\ref{eq:severity_difference}, and the absolute severity shift defined in Eq.~\ref{eq:absolute_severity_shift}. Severity preservation was summarized overall and by speaker group as the proportion of speakers with unchanged severity score, a one-level shift, and a shift of two or more levels. Because the severity score was ordinal and paired within speaker, original and anonymized severity scores were compared with paired non-parametric tests, using the Wilcoxon signed-rank test when its assumptions were appropriate and an exact sign test if the distribution of paired differences made the signed-rank test unsuitable. Binary pathological status was defined as severity score greater than 0, and preservation of pathological vs non-pathological classification was evaluated with paired binary agreement summaries and McNemar's test when discordant counts were sufficient; otherwise, exact paired-binomial inference was used. These analyses were reported for the full cohort and, where sample size permitted, separately for the six speaker groups.

Listener-language and domain-expertise analyses used listener-by-group cells as the analysis unit. Native-language effects compared native and non-native German listeners on zero-shot accuracy, few-shot accuracy, and QDS. Domain-expertise effects compared listeners with at least 8 years of specialized experience in signal-level speech processing or clinical voice and speech assessment against the remaining listeners. Because these comparisons involved independent listener groups and small samples, two-sided Mann-Whitney U tests \cite{mann1947test,mannwit} were used. Standardized effect sizes were reported as Cohen's $d$ \cite{cohen2013statistical}, with positive values defined to favor the first category in the plotted contrast, namely native over non-native listeners or experts over non-experts.

Speaker-sex analyses used per-speaker outcomes averaged across the 10 listeners. Female and male speakers were compared for zero-shot accuracy, few-shot accuracy, and QDS using two-sided Mann-Whitney U tests across the full cohort and within each speaker group. Speaker-age associations were evaluated with Spearman rank correlations using per-speaker outcomes. Age correlations were computed across the full cohort and again after restricting the analysis to the four adult groups, namely Dysarthria, Dysglossia, Dysphonia, and adult controls.

Associations between perceptual outcomes and computational reference metrics were analyzed at the pathology-group level for the four pathology groups with available computational measurements: CLP, Dysarthria, Dysglossia, and Dysphonia. Pearson correlations were computed between group-level perceptual measures, including zero-shot accuracy, few-shot accuracy, and QDS, and computational metrics, including McAdams EER, post-anonymization AUROC, and $\Delta\mathrm{AUROC}$. Because these analyses used only four group means, they were interpreted as descriptive evidence of alignment or dissociation rather than as high-powered inferential tests. Multi-axis signatures and composite visualizations were constructed from the same group-level summaries using z-normalization or min-max rescaling only for visualization, as indicated in the corresponding figure captions.





\section*{Data availability}

The dataset used in this study is internal data of patients of the University Hospital Erlangen and is not publicly available due to patient privacy regulations. A reasonable request to the corresponding author is required for accessing the data on-site at the University Hospital Erlangen in Erlangen, Germany.


\section*{Code availability}

To encourage transparency and facilitate future research, we have publicly released our complete source code at \url{https://github.com/tayebiarasteh/perceptual}. The code is implemented in Python v3.10 and leverages the PyTorch v2.1 framework for all deep learning operations. All statistical analyses were performed using the NumPy v1.22, pandas v1.4, SciPy v1.7, and statsmodels v0.14 libraries.


\section*{Acknowledgements}

No funding was received for this study.


\section*{Author contributions}

The formal analysis was conducted by STA and AM. The original draft was written by STA. The software was developed by STA. The perceptual tests were designed by STA and AM. The listening tests were performed by TN, SA, LB, TG, HH, MS, ML, TA, MP, and EN. Evaluation and statistical analysis were performed by STA. Datasets were provided by EN, MS, SHY, and AM. STA cleaned, organized, and preprocessed the data. STA, TN, and MS provided clinical expertise. MS provided phoniatrics expertise. STA, SA, TN, LB, MP, TA, PAPT, ML, TG, EN, SHY, and AM, provided technical expertise. STA and AM designed the study. All authors read the manuscript, contributed to the editing, and agreed to the submission of this paper.


\section*{Declaration of interests}

STA is on the editorial board of Communications Medicine and of European Radiology Experimental, and on the trainee editorial board of Radiology: Artificial Intelligence. ML is employed by Generali Deutschland Services GmbH, Germany, and is on the editorial board of European Radiology Experimental. AM is an associate editor at IEEE Transactions on Medical Imaging. The other authors do not have any competing interests to disclose.


\bibliographystyle{splncs04}
\bibliography{bibliography}

@article{arasteh2024addressing,
  title={Addressing challenges in speaker anonymization to maintain utility while ensuring privacy of pathological speech},
  author={Tayebi Arasteh, Soroosh and Arias-Vergara, Tom{\'a}s and P{\'e}rez-Toro, Paula Andrea and Weise, Tobias and Packh{\"a}user, Kai and Schuster, Maria and Noeth, Elmar and Maier, Andreas and Yang, Seung Hee},
  journal={Communications Medicine},
  volume={4},
  number={1},
  pages={182},
  year={2024},
  publisher={Nature Publishing Group UK London}
}

@article{arasteh2023effect,
  title={The effect of speech pathology on automatic speaker verification: a large-scale study},
  author={Tayebi Arasteh, Soroosh and Weise, Tobias and Schuster, Maria and Noeth, Elmar and Maier, Andreas and Yang, Seung Hee},
  journal={Scientific Reports},
  volume={13},
  number={1},
  pages={20476},
  year={2023},
  publisher={Nature Publishing Group UK London}
}

@article{maier2009peaks,
  title={PEAKS--A system for the automatic evaluation of voice and speech disorders},
  author={Maier, Andreas and Haderlein, Tino and Eysholdt, Ulrich and Rosanowski, Frank and Batliner, Anton and Schuster, Maria and N{\"o}th, Elmar},
  journal={Speech Communication},
  volume={51},
  number={5},
  pages={425--437},
  year={2009},
  publisher={Elsevier}
}

@article{harding1996characteristics,
author = {Harding, Anne and Grunwell, Pamela},
title = {Characteristics of cleft palate speech},
journal = {International Journal of Language \& Communication Disorders},
volume = {31},
number = {4},
pages = {331-357},
url = {https://onlinelibrary.wiley.com/doi/abs/10.3109/13682829609031326},
year = {1996}
}

@article{maier2006fully,
  title={Fully automatic assessment of speech of children with cleft lip and palate},
  author={Maier, Andreas and N{\"o}th, Elmar and Batliner, Anton and Nkenke, Emeka and Schuster, Maria},
  journal={Informatica},
  volume={30},
  number={4},
  year={2006}
}

@article{millard2001different,
  title={Different cleft conditions, facial appearance, and speech: relationship to psychological variables},
  author={Millard, Tom and Richman, Lynn C},
  journal={The Cleft palate-craniofacial journal},
  volume={38},
  number={1},
  pages={68--75},
  year={2001},
  publisher={SAGE Publications Sage CA: Los Angeles, CA}
}

@article{hirose1986pathophysiology,
    author = {Hirose, Hajime},
    title = {Pathophysiology of Motor Speech Disorders (Dysarthria)},
    journal = {Folia Phoniatrica et Logopaedica},
    volume = {38},
    number = {2-4},
    pages = {61-88},
    year = {2009},
    url = {https://doi.org/10.1159/000265824},
}

@article{schroter2005rehabilitation,
  title={Rehabilitation of impaired speech function (dysarthria, dysglossia)},
  author={Schr{\"o}ter-Morasch, Heidrun and Ziegler, Wolfram},
  journal={GMS current topics in otorhinolaryngology, head and neck surgery},
  volume={4},
  pages={Doc15},
  year={2005}
}

@article{sama2001clinical,
author = {Sama, A. and Carding, P. N. and Price, S. and Kelly, P. and Wilson, J. A.},
title = {The Clinical Features of Functional Dysphonia},
journal = {The Laryngoscope},
volume = {111},
number = {3},
pages = {458-463},
url = {https://onlinelibrary.wiley.com/doi/abs/10.1097/00005537-200103000-00015},
year = {2001}
}

@book{fox2002plakss,
  title={PLAKSS: psycholinguistische Analyse kindlicher Sprechst{\"o}rungen},
  author={Fox-Boyer, A.},
  url={https://books.google.de/books?id=A1cwpwAACAAJ},
  year={2002},
  publisher={Swets Test Services}
}

@inproceedings{patino2021speaker,
  title     = {{Speaker Anonymisation Using the McAdams Coefficient}},
  author    = {Jose Patino and Natalia Tomashenko and Massimiliano Todisco and Andreas Nautsch and Nicholas Evans},
  year      = {2021},
  booktitle = {{Interspeech 2021}},
  pages     = {1099--1103},
  doi       = {10.21437/Interspeech.2021-1070},
  issn      = {2958-1796},
}

@mastersthesis {mcadams1984spectral,
	title = {Spectral Fusion, Spectral Parsing and the Formation of Auditory Images},
	volume = {Ph.D.},
	year = {1984},
	month = {05/1984},
	school = {Stanford University},
	address = {Stanford, California},
	url = {https://ccrma.stanford.edu/files/papers/stanm22.pdf},
	author = {S. McAdams}
}

@book{little2020common,
  title={Common European framework of reference for languages: Learning, teaching, assessment},
  author={The Council of Europe,.},
  year={2001},
  publisher={Cambridge University Press}
}

@article{turing1950computing,
    author = {Turing, A. M.},
    title = {I.—COMPUTING MACHINERY AND INTELLIGENCE},
    journal = {Mind},
    volume = {LIX},
    number = {236},
    pages = {433-460},
    year = {1950},
    issn = {0026-4423},
    doi = {10.1093/mind/LIX.236.433},
    url = {https://doi.org/10.1093/mind/LIX.236.433},
}

@inproceedings{kingma2015adam,
  author    = {Kingma, Diederik P. and Ba, Jimmy},
  title     = {Adam: A Method for Stochastic Optimization},
  booktitle = {International Conference on Learning Representations (ICLR)},
  year      = {2015},
  url       = {https://arxiv.org/abs/1412.6980}
}

@INPROCEEDINGS{panayotov2015librispeech,
  author={Panayotov, Vassil and Chen, Guoguo and Povey, Daniel and Khudanpur, Sanjeev},
  booktitle={2015 IEEE International Conference on Acoustics, Speech and Signal Processing (ICASSP)}, 
  title={Librispeech: An ASR corpus based on public domain audio books}, 
  year={2015},
  volume={},
  number={},
  pages={5206-5210},
  doi={10.1109/ICASSP.2015.7178964}}

@inproceedings{wan2018generalized,
  title={Generalized end-to-end loss for speaker verification},
  author={Wan, Li and Wang, Quan and Papir, Alan and Moreno, Ignacio Lopez},
  booktitle={2018 IEEE International Conference on Acoustics, Speech and Signal Processing (ICASSP)},
  pages={4879--4883},
  year={2018},
  organization={IEEE}
}

@misc{arasteh2022empirical,
      title={An Empirical Study on Text-Independent Speaker Verification based on the GE2E Method}, 
      author={Soroosh Tayebi Arasteh},
      year={2022},
      eprint={2011.04896},
      archivePrefix={arXiv},
      primaryClass={eess.AS},
      url={https://arxiv.org/abs/2011.04896}, 
}

@INPROCEEDINGS{deng2009imagenet,
  author={Deng, Jia and Dong, Wei and Socher, Richard and Li, Li-Jia and Kai Li and Li Fei-Fei},
  booktitle={2009 IEEE Conference on Computer Vision and Pattern Recognition}, 
  title={ImageNet: A large-scale hierarchical image database}, 
  year={2009},
  volume={},
  number={},
  pages={248-255},
  doi={10.1109/CVPR.2009.5206848}}

@INPROCEEDINGS {he2016deep,
author = {He, Kaiming and Zhang, Xiangyu and Ren, Shaoqing and Sun, Jian },
booktitle = { 2016 IEEE Conference on Computer Vision and Pattern Recognition (CVPR) },
title = {{ Deep Residual Learning for Image Recognition }},
year = {2016},
volume = {},
pages = {770-778},
url = {https://doi.ieeecomputersociety.org/10.1109/CVPR.2016.90},
}

@article{
anova_main,
author = {Martin G. Larson },
title = {Analysis of Variance},
journal = {Circulation},
volume = {117},
number = {1},
pages = {115-121},
year = {2008},
URL = {https://www.ahajournals.org/doi/abs/10.1161/CIRCULATIONAHA.107.654335},
}

@article{
repeatedmeasures,
author = {Lisa M. Sullivan },
title = {Repeated Measures},
journal = {Circulation},
volume = {117},
number = {9},
pages = {1238-1243},
year = {2008},
URL = {https://www.ahajournals.org/doi/abs/10.1161/CIRCULATIONAHA.107.654350},
}

@book {BasicandAdvancedStatisticalTests,
      author = "Amanda Ross",
      title = "Basic and Advanced Statistical Tests: Writing Results Sections and Creating Tables and Figures",
      year = "2019",
      publisher = "Brill",
      address = "Leiden, The Netherlands",
      doi = "10.1007/978-94-6351-086-8",
      url = "https://brill.com/view/title/37997"
}

@inbook {OneWayANOVA,
      author = "Amanda Ross and Victor L. Willson",
      title = "One-Way ANOVA",
      booktitle = "",
      year = "2017",
      address = "Leiden, The Netherlands",
      isbn = "9789463510868",
      pages=      "21 - 24",
      url = "https://brill.com/view/book/9789463510868/BP000006.xml"
}

@article{mann1947test,
  title={On a test of whether one of two random variables is stochastically larger than the other},
  author={Mann, Henry B and Whitney, Donald R},
  journal={The annals of mathematical statistics},
  pages={50--60},
  year={1947},
  publisher={JSTOR}
}

@inbook{mannwit,
author = {McKnight, Patrick E. and Najab, Julius},
publisher = {John Wiley \& Sons, Ltd},
isbn = {9780470479216},
title = {Mann-Whitney U Test},
booktitle = {The Corsini Encyclopedia of Psychology},
chapter = {},
pages = {1-1},
url = {https://onlinelibrary.wiley.com/doi/abs/10.1002/9780470479216.corpsy0524},
year = {2010},
}

@article{benjamini1995controlling,
  title={Controlling the false discovery rate: a practical and powerful approach to multiple testing},
  author={Benjamini, Yoav and Hochberg, Yosef},
  journal={Journal of the Royal statistical society: series B (Methodological)},
  volume={57},
  number={1},
  pages={289--300},
  year={1995},
  publisher={Wiley Online Library}
}

@article{riedhammer2023medical,
  title={Medical Speech Processing for Diagnosis and Monitoring: Clinical Use Cases},
  author={Riedhammer, Korbinian and Baumann, Ilja and Bayerl, Sebastian P and Bocklet, Tobias and Braun, Franziska and Wagner, Dominik},
  year={2023}
}

@inproceedings{pappagari20_interspeech,
  title     = {{Using State of the Art Speaker Recognition and Natural Language Processing Technologies to Detect Alzheimer’s Disease and Assess its Severity}},
  author    = {Raghavendra Pappagari and Jaejin Cho and Laureano Moro-Velázquez and Najim Dehak},
  year      = {2020},
  booktitle = {{Interspeech 2020}},
  pages     = {2177--2181},
  doi       = {10.21437/Interspeech.2020-2587},
  issn      = {2958-1796},
}

@article{bayerl2022can,
  title={What can speech and language tell us about the working alliance in psychotherapy},
  author={Bayerl, Sebastian P and Roccabruna, Gabriel and Chowdhury, Shammur Absar and Ciulli, Tommaso and Danieli, Morena and Riedhammer, Korbinian and Riccardi, Giuseppe},
  journal={arXiv preprint arXiv:2206.08835},
  year={2022}
}

@article{strimbu2010biomarkers,
  title={What are biomarkers?},
  author={Strimbu, Kyle and Tavel, Jorge A},
  journal={Current Opinion in HIV and AIDS},
  volume={5},
  number={6},
  pages={463--466},
  year={2010},
  publisher={LWW}
}

@article{califf2018biomarker,
  title={Biomarker definitions and their applications},
  author={Califf, Robert M},
  journal={Experimental biology and medicine},
  volume={243},
  number={3},
  pages={213--221},
  year={2018},
  publisher={SAGE Publications Sage UK: London, England}
}

@article{
speechbiomar,
author = {Vikram Ramanarayanan  and Adam C. Lammert  and Hannah P. Rowe  and Thomas F. Quatieri  and Jordan R. Green },
title = {Speech as a Biomarker: Opportunities, Interpretability, and Challenges},
journal = {Perspectives of the ASHA Special Interest Groups},
volume = {7},
number = {1},
pages  = {276-283},
year = {2022},
doi = {10.1044/2021\_PERSP-21-00174},
}

@incollection{kroger2019privacy,
  title={Privacy implications of voice and speech analysis--information disclosure by inference},
  author={Kr{\"o}ger, Jacob Leon and Lutz, Otto Hans-Martin and Raschke, Philip},
  booktitle={IFIP International Summer School on Privacy and Identity Management},
  pages={242--258},
  year={2019},
  publisher={Springer}
}

@article{tomashenko2022voiceprivacy,
title = {The VoicePrivacy 2020 Challenge: Results and findings},
journal = {Computer Speech \& Language},
volume = {74},
pages = {101362},
year = {2022},
url = {https://www.sciencedirect.com/science/article/pii/S0885230822000080},
author = {Natalia Tomashenko and Xin Wang and Emmanuel Vincent and Jose Patino and Brij Mohan Lal Srivastava and Paul-Gauthier Noé and Andreas Nautsch and Nicholas Evans and Junichi Yamagishi and Benjamin O’Brien and Anaïs Chanclu and Jean-François Bonastre and Massimiliano Todisco and Mohamed Maouche},
}

@misc{tomashenko2024voiceprivacy,
      title={The VoicePrivacy 2024 Challenge Evaluation Plan}, 
      author={Natalia Tomashenko and Xiaoxiao Miao and Pierre Champion and Sarina Meyer and Xin Wang and Emmanuel Vincent and Michele Panariello and Nicholas Evans and Junichi Yamagishi and Massimiliano Todisco},
      year={2024},
      eprint={2404.02677},
      archivePrefix={arXiv},
      primaryClass={eess.AS},
      url={https://arxiv.org/abs/2404.02677}, 
}

@misc{tomashenko2022voiceprivacy2022,
      title={The VoicePrivacy 2022 Challenge Evaluation Plan}, 
      author={Natalia Tomashenko and Xin Wang and Xiaoxiao Miao and Hubert Nourtel and Pierre Champion and Massimiliano Todisco and Emmanuel Vincent and Nicholas Evans and Junichi Yamagishi and Jean-François Bonastre},
      year={2022},
      eprint={2203.12468},
      archivePrefix={arXiv},
      primaryClass={eess.AS},
      url={https://arxiv.org/abs/2203.12468}, 
}

@INPROCEEDINGS{srivastava2022privacy,
  author={Lal Srivastava, Brij Mohan and Vauquier, Nathalie and Sahidullah, Md and Bellet, Aurélien and Tommasi, Marc and Vincent, Emmanuel},
  booktitle={ICASSP 2020 - IEEE International Conference on Acoustics, Speech and Signal Processing (ICASSP)}, 
  title={Evaluating Voice Conversion-Based Privacy Protection against Informed Attackers}, 
  year={2020},
  volume={},
  number={},
  pages={2802-2806},
  doi={10.1109/ICASSP40776.2020.9053868}
  }

@inproceedings{fang2019speaker,
  title     = {{Speaker Anonymization Using X-vector and Neural Waveform Models}},
  author    = {Fuming Fang and Xin Wang and Junichi Yamagishi and Isao Echizen and Massimiliano Todisco and Nicholas Evans and Jean-Francois Bonastre},
  year      = {2019},
  booktitle = {{10th ISCA Workshop on Speech Synthesis (SSW 10)}},
  pages     = {155--160},
  doi       = {10.21437/SSW.2019-28},
}

@inproceedings{khamsehashari22_spsc,
  title     = {{Voice Privacy - leveraging multi-scale blocks with ECAPA-TDNN SE-Res2NeXt extension for speaker anonymization}},
  author    = {Razieh Khamsehashari and Yamini Sinha and Jan Hintz and Suhita Ghosh and Tim Polzehl and Carlos Franzreb and Sebastian Stober and Ingo Siegert},
  year      = {2022},
  booktitle = {{2nd Symposium on Security and Privacy in Speech Communication}},
  pages     = {43--48},
  doi       = {10.21437/SPSC.2022-8},
}

@INPROCEEDINGS{Snyderxvector,
  author={Snyder, David and Garcia-Romero, Daniel and Sell, Gregory and Povey, Daniel and Khudanpur, Sanjeev},
  booktitle={2018 IEEE International Conference on Acoustics, Speech and Signal Processing (ICASSP)}, 
  title={X-Vectors: Robust DNN Embeddings for Speaker Recognition}, 
  year={2018},
  volume={},
  number={},
  pages={5329-5333},
  doi={10.1109/ICASSP.2018.8461375}}

@article{NAUTSCH2019441,
title = {Preserving privacy in speaker and speech characterisation},
journal = {Computer Speech \& Language},
volume = {58},
pages = {441-480},
year = {2019},
issn = {0885-2308},
doi = {https://doi.org/10.1016/j.csl.2019.06.001},
url = {https://www.sciencedirect.com/science/article/pii/S0885230818303875},
author = {Andreas Nautsch and Abelino Jiménez and Amos Treiber and Jascha Kolberg and Catherine Jasserand and Els Kindt and Héctor Delgado and Massimiliano Todisco and Mohamed Amine Hmani and Aymen Mtibaa and Mohammed Ahmed Abdelraheem and Alberto Abad and Francisco Teixeira and Driss Matrouf and Marta Gomez-Barrero and Dijana Petrovska-Delacrétaz and Gérard Chollet and Nicholas Evans and Thomas Schneider and Jean-François Bonastre and Bhiksha Raj and Isabel Trancoso and Christoph Busch},
}

@article{tayebi2026privacy,
  title={A privacy stack for speech-based AI in digital health},
  author={Tayebi Arasteh, Soroosh},
  journal={npj Digital Public Health},
  volume={1},
  number={1},
  pages={12},
  year={2026},
  publisher={Nature Publishing Group UK London}
}

@inproceedings{srivastava20_interspeech,
  title     = {{Design Choices for X-Vector Based Speaker Anonymization}},
  author    = {Brij Mohan Lal Srivastava and N. Tomashenko and Xin Wang and Emmanuel Vincent and Junichi Yamagishi and Mohamed Maouche and Aurélien Bellet and Marc Tommasi},
  year      = {2020},
  booktitle = {{Interspeech 2020}},
  pages     = {1713--1717},
  doi       = {10.21437/Interspeech.2020-2692},
  issn      = {2958-1796},
}

@inproceedings{siegert2024user,
    title = "User Perspective on Anonymity in Voice Assistants {--} A comparison between {G}ermany and {F}inland",
    author = {Siegert, Ingo  and
      Rech, Silas  and
      B{\"a}ckstr{\"o}m, Tom  and
      Haase, Matthias},
    editor = "Siegert, Ingo  and
      Choukri, Khalid",
    booktitle = "Proceedings of the Workshop on Legal and Ethical Issues in Human Language Technologies @ LREC-COLING 2024",
    month = may,
    year = "2024",
    address = "Torino, Italia",
    publisher = "ELRA and ICCL",
    url = "https://aclanthology.org/2024.legal-1.11/",
    pages = "73--78",
}

@article{
Kluinpercept,
author = {Karen J. Kluin  and Norman L. Foster  and Stanley Berent  and Sid Gilman },
title = {Perceptual analysis of speech disorders in progressive supranuclear palsy},
journal = {Neurology},
volume = {43},
number = {3\_part\_1},
pages = {563-563},
year = {1993},
URL = {https://www.neurology.org/doi/abs/10.1212/WNL.43.3_Part_1.563},
}

@article{sachin2008clinical,
  title={Clinical speech impairment in Parkinson's disease, progressive supranuclear palsy, and multiple system atrophy},
  author={Sachin, S and Shukla, Garima and Goyal, Vaibhav and Singh, Shivangi and Aggarwal, Vijay and Behari, Madhuri and others},
  journal={Neurology India},
  volume={56},
  number={2},
  pages={122--126},
  year={2008},
  publisher={Medknow}
}

@article{
Pernonpercep,
author = {Michaela Pernon  and Frédéric Assal  and Ina Kodrasi  and Marina Laganaro },
title = {Perceptual Classification of Motor Speech Disorders: The Role of Severity, Speech Task, and Listener's Expertise},
journal = {Journal of Speech, Language, and Hearing Research},
volume = {65},
number = {8},
pages  = {2727-2747},
year = {2022},
URL = {https://pubs.asha.org/doi/abs/10.1044/2022_JSLHR-21-00519},
}

@inproceedings{tayebiarasteh23_interspeech,
  title     = {{Federated Learning for Secure Development of AI Models for Parkinson’s Disease Detection Using Speech from Different Languages}},
  author    = {Soroosh {Tayebi Arasteh} and Cristian David Ríos-Urrego and Elmar Nöth and Andreas Maier and Seung Hee Yang and Jan Rusz and Juan Rafael Orozco-Arroyave},
  year      = {2023},
  booktitle = {{Interspeech 2023}},
  pages     = {5003--5007},
  doi       = {10.21437/Interspeech.2023-2108},
  issn      = {2958-1796},
}

@article{mohammadi2026differential,
  title={Differential privacy for medical deep learning: methods, tradeoffs, and deployment implications},
  author={Mohammadi, Marziyeh and Vejdanihemmat, Mohsen and Lotfinia, Mahshad and Rusu, Mirabela and Truhn, Daniel and Maier, Andreas and Tayebi Arasteh, Soroosh},
  journal={npj Digital Medicine},
  year={2026},
  volume={9},
  pages={93},
  publisher={Nature Publishing Group UK London}
}

@article{kent1996hearing,
  title={Hearing and believing: Some limits to the auditory-perceptual assessment of speech and voice disorders},
  author={Kent, Ray D},
  journal={American Journal of Speech-Language Pathology},
  volume={5},
  number={3},
  pages={7--23},
  year={1996},
  publisher={American Speech-Language-Hearing Association Rockville, MD}
}

@misc{champion2024anonymizingspeechevaluatingdesigning,
      title={Anonymizing Speech: Evaluating and Designing Speaker Anonymization Techniques}, 
      author={Pierre Champion},
      year={2024},
      eprint={2308.04455},
      archivePrefix={arXiv},
      primaryClass={cs.CR},
      url={https://arxiv.org/abs/2308.04455}, 
}

@article{hsu2021hubert,
author = {Hsu, Wei-Ning and Bolte, Benjamin and Tsai, Yao-Hung Hubert and Lakhotia, Kushal and Salakhutdinov, Ruslan and Mohamed, Abdelrahman},
title = {HuBERT: Self-Supervised Speech Representation Learning by Masked Prediction of Hidden Units},
year = {2021},
issue_date = {2021},
publisher = {IEEE Press},
volume = {29},
issn = {2329-9290},
url = {https://doi.org/10.1109/TASLP.2021.3122291},
journal = {IEEE/ACM Trans. Audio, Speech and Lang. Proc.},
pages = {3451–3460},
numpages = {10}
}

@article{
defossez2023encodec,
title={High Fidelity Neural Audio Compression},
author={Alexandre D{\'e}fossez and Jade Copet and Gabriel Synnaeve and Yossi Adi},
journal={Transactions on Machine Learning Research},
issn={2835-8856},
year={2023},
url={https://openreview.net/forum?id=ivCd8z8zR2},
note={Featured Certification, Reproducibility Certification}
}

@INPROCEEDINGS{Panariellospeak,
  author={Panariello, Michele and Nespoli, Francesco and Todisco, Massimiliano and Evans, Nicholas},
  booktitle={ICASSP 2024 - 2024 IEEE International Conference on Acoustics, Speech and Signal Processing (ICASSP)}, 
  title={Speaker Anonymization Using Neural Audio Codec Language Models}, 
  year={2024},
  volume={},
  number={},
  pages={4725-4729},
  doi={10.1109/ICASSP48485.2024.10447871}}

@article{Borsosnac,
author = {Borsos, Zal\'{a}n and Marinier, Rapha\"{e}l and Vincent, Damien and Kharitonov, Eugene and Pietquin, Olivier and Sharifi, Matt and Roblek, Dominik and Teboul, Olivier and Grangier, David and Tagliasacchi, Marco and Zeghidour, Neil},
title = {AudioLM: A Language Modeling Approach to Audio Generation},
year = {2023},
issue_date = {2023},
publisher = {IEEE Press},
volume = {31},
url = {https://doi.org/10.1109/TASLP.2023.3288409},
journal = {IEEE/ACM Trans. Audio, Speech and Lang. Proc.},
pages = {2523–2533},
numpages = {11}
}

@ARTICLE{Chennac,
  author={Chen, Sanyuan and Wang, Chengyi and Wu, Yu and Zhang, Ziqiang and Zhou, Long and Liu, Shujie and Chen, Zhuo and Liu, Yanqing and Wang, Huaming and Li, Jinyu and He, Lei and Zhao, Sheng and Wei, Furu},
  journal={IEEE Transactions on Audio, Speech and Language Processing}, 
  title={Neural Codec Language Models are Zero-Shot Text to Speech Synthesizers}, 
  year={2025},
  volume={33},
  number={},
  pages={705-718},
  doi={10.1109/TASLPRO.2025.3530270}}

@inbook{fastspeech,
author = {Ren, Yi and Ruan, Yangjun and Tan, Xu and Qin, Tao and Zhao, Sheng and Zhao, Zhou and Liu, Tie-Yan},
title = {FastSpeech: fast, robust and controllable text to speech},
year = {2019},
publisher = {Curran Associates Inc.},
address = {Red Hook, NY, USA},
booktitle = {Proceedings of the 33rd International Conference on Neural Information Processing Systems},
articleno = {285},
numpages = {10}
}

@inproceedings{
ren2020fastspeech2,
title={FastSpeech 2: Fast and High-Quality End-to-End Text to Speech},
author={Yi Ren and Chenxu Hu and Xu Tan and Tao Qin and Sheng Zhao and Zhou Zhao and Tie-Yan Liu},
booktitle={International Conference on Learning Representations},
year={2021},
url={https://openreview.net/forum?id=piLPYqxtWuA}
}

@inproceedings{lux23_blizzard,
  title     = {{The IMS Toucan System for the Blizzard Challenge 2023}},
  author    = {Florian Lux and Julia Koch and Sarina Meyer and Thomas Bott and Nadja Schauffler and Pavel Denisov and Antje Schweitzer and Ngoc Thang Vu},
  year      = {2023},
  booktitle = {{18th Blizzard Challenge Workshop}},
  pages     = {40--45},
  doi       = {10.21437/Blizzard.2023-4},
}

@INPROCEEDINGS{10096607,
  author={Meyer, Sarina and Lux, Florian and Koch, Julia and Denisov, Pavel and Tilli, Pascal and Vu, Ngoc Thang},
  booktitle={ICASSP 2023 - 2023 IEEE International Conference on Acoustics, Speech and Signal Processing (ICASSP)}, 
  title={Prosody Is Not Identity: A Speaker Anonymization Approach Using Prosody Cloning}, 
  year={2023},
  volume={},
  number={},
  pages={1-5},
  doi={10.1109/ICASSP49357.2023.10096607}}

@inproceedings{meyer2023anonymizing,
  title={Anonymizing speech with generative adversarial networks to preserve speaker privacy},
  author={Meyer, Sarina and Tilli, Pascal and Denisov, Pavel and Lux, Florian and Koch, Julia and Vu, Ngoc Thang},
  booktitle={2022 IEEE Spoken Language Technology Workshop (SLT)},
  pages={912--919},
  year={2023},
  organization={IEEE}
}

@inproceedings{kong2020hifigan,
 author = {Kong, Jungil and Kim, Jaehyeon and Bae, Jaekyoung},
 booktitle = {Advances in Neural Information Processing Systems},
 editor = {H. Larochelle and M. Ranzato and R. Hadsell and M.F. Balcan and H. Lin},
 pages = {17022--17033},
 title = {HiFi-GAN: Generative Adversarial Networks for Efficient and High Fidelity Speech Synthesis},
 url = {https://proceedings.neurips.cc/paper_files/paper/2020/file/c5d736809766d46260d816d8dbc9eb44-Paper.pdf},
 volume = {33},
 year = {2020}
}

@article{ganmain,
author = {Goodfellow, Ian and Pouget-Abadie, Jean and Mirza, Mehdi and Xu, Bing and Warde-Farley, David and Ozair, Sherjil and Courville, Aaron and Bengio, Yoshua},
title = {Generative adversarial networks},
year = {2020},
issue_date = {November 2020},
publisher = {Association for Computing Machinery},
address = {New York, NY, USA},
volume = {63},
number = {11},
issn = {0001-0782},
url = {https://doi.org/10.1145/3422622},
doi = {10.1145/3422622},
journal = {Commun. ACM},
month = oct,
pages = {139–144},
numpages = {6}
}

@inproceedings{Wassersteingan,
author = {Arjovsky, Martin and Chintala, Soumith and Bottou, L\'{e}on},
title = {Wasserstein generative adversarial networks},
year = {2017},
publisher = {JMLR.org},
booktitle = {Proceedings of the 34th International Conference on Machine Learning - Volume 70},
pages = {214–223},
numpages = {10},
location = {Sydney, NSW, Australia},
series = {ICML'17}
}

@book{cohen2013statistical,
  title={Statistical power analysis for the behavioral sciences},
  author={Cohen, Jacob},
  year={2013},
  publisher={routledge}
}

@article{cohen1968weighted,
  title={Weighted kappa: Nominal scale agreement provision for scaled disagreement or partial credit.},
  author={Cohen, Jacob},
  journal={Psychological bulletin},
  volume={70},
  number={4},
  pages={213},
  year={1968},
  publisher={American Psychological Association}
}

@article{spearman1961proof,
  title={The proof and measurement of association between two things.},
  author={Spearman, Charles},
  year={1961},
  publisher={Appleton-Century-Crofts}
}

@article{tayebi2025differential,
  title={Differential privacy enables fair and accurate AI-based analysis of speech disorders while protecting patient data},
  author={Tayebi Arasteh, Soroosh and Lotfinia, Mahshad and Perez-Toro, Paula Andrea and Arias-Vergara, Tomas and Ranji, Mahtab and Orozco-Arroyave, Juan Rafael and Schuster, Maria and Maier, Andreas and Yang, Seung Hee},
  journal={npj Artificial Intelligence},
  volume={1},
  number={1},
  pages={37},
  year={2025},
  publisher={Nature Publishing Group UK London}
}

@article{kinnunen2010overview,
  title={An overview of text-independent speaker recognition: From features to supervectors},
  author={Kinnunen, Tomi and Li, Haizhou},
  journal={Speech communication},
  volume={52},
  number={1},
  pages={12--40},
  year={2010},
  publisher={Elsevier}
}

@article{hansen2015speaker,
  title={Speaker recognition by machines and humans: A tutorial review},
  author={Hansen, John HL and Hasan, Taufiq},
  journal={IEEE Signal processing magazine},
  volume={32},
  number={6},
  pages={74--99},
  year={2015},
  publisher={IEEE}
}

@INPROCEEDINGS{7472729,
  author={Hashimoto, Kei and Yamagishi, Junichi and Echizen, Isao},
  booktitle={2016 IEEE International Conference on Acoustics, Speech and Signal Processing (ICASSP)}, 
  title={Privacy-preserving sound to degrade automatic speaker verification performance}, 
  year={2016},
  volume={},
  number={},
  pages={5500-5504},
}

@article{schmidhuber1997long,
  title={Long short-term memory},
  author={Schmidhuber, J{\"u}rgen and Hochreiter, Sepp and others},
  journal={Neural Comput},
  volume={9},
  number={8},
  pages={1735--1780},
  year={1997}
}

@inproceedings{sak2014long,
  title={Long short-term memory recurrent neural network architectures for large scale acoustic modeling.},
  author={Sak, Hasim and Senior, Andrew W and Beaufays, Fran{\c{c}}oise and others},
  booktitle={Interspeech},
  volume={2014},
  pages={338--342},
  year={2014}
}

@article{console46plantronics,
  title={Plantronics inc.},
  journal={Santa cruz, CA, USA},
  url={https://www.poly.com}
}

@article{logitechk,
  title={Logitech International S.A.},
  journal={Lausanne, Switzerland},
  url={https://www.logitech.com/}
}

@article{ramirez2007voice,
  title={Voice activity detection. fundamentals and speech recognition system robustness},
  author={Ramirez, Javier and G{\'o}rriz, Juan Manuel and Segura, Jos{\'e} Carlos},
  journal={Robust speech recognition and understanding},
  volume={6},
  number={9},
  pages={1--22},
  year={2007},
  publisher={Vienna, Austria}
}

@inproceedings{prabhavalkar2015automatic,
  title={Automatic gain control and multi-style training for robust small-footprint keyword spotting with deep neural networks},
  author={Prabhavalkar, Rohit and Alvarez, Raziel and Parada, Carolina and Nakkiran, Preetum and Sainath, Tara N},
  booktitle={2015 IEEE International Conference on Acoustics, Speech and Signal Processing (ICASSP)},
  pages={4704--4708},
  year={2015},
  organization={IEEE}
}

@article{gustafsson2002determining,
  title={Determining the initial states in forward-backward filtering},
  author={Gustafsson, Fredrik},
  journal={IEEE Transactions on signal processing},
  volume={44},
  number={4},
  pages={988--992},
  year={2002},
  publisher={IEEE}
}

@article{chlasta2019automated,
  title={Automated speech-based screening of depression using deep convolutional neural networks},
  author={Chlasta, Karol and Wo{\l}k, Krzysztof and Krejtz, Izabela},
  journal={Procedia Computer Science},
  volume={164},
  pages={618--628},
  year={2019},
  publisher={Elsevier}
}

@article{muzammel2020audvowelconsnet,
  title={AudVowelConsNet: A phoneme-level based deep CNN architecture for clinical depression diagnosis},
  author={Muzammel, Muhammad and Salam, Hanan and Hoffmann, Yann and Chetouani, Mohamed and Othmani, Alice},
  journal={Machine Learning with Applications},
  volume={2},
  pages={100005},
  year={2020},
  publisher={Elsevier}
}

@article{darley1969differential,
  title={Differential diagnostic patterns of dysarthria},
  author={Darley, Frederic L and Aronson, Arnold E and Brown, Joe R},
  journal={Journal of speech and hearing research},
  volume={12},
  number={2},
  pages={246--269},
  year={1969},
  publisher={American Speech-Language-Hearing Association}
}

@inproceedings{kim2010dysarthria,
  title={Dysarthric speech database for universal access research.},
  author={Kim, Heejin and Hasegawa-Johnson, Mark and Perlman, Adrienne and Gunderson, Jon R and Huang, Thomas S and Watkin, Kenneth L and Frame, Simone and others},
  booktitle={Interspeech},
  volume={2008},
  pages={1741--1744},
  year={2008}
}

@misc{hirano1981clinical,
  title={Clinical examination of voice by Minoru Hirano},
  author={Hirano, Minoru and McCormick, Karen R},
  year={1986},
  publisher={Acoustical Society of America}
}

@article{adank2010comprehension,
  title={Comprehension of familiar and unfamiliar native accents under adverse listening conditions.},
  author={Adank, Patti and Evans, Bronwen G and Stuart-Smith, Jane and Scott, Sophie K},
  journal={Journal of Experimental Psychology: Human perception and performance},
  volume={35},
  number={2},
  pages={520},
  year={2009},
  publisher={American Psychological Association}
}

@article{bradlow2008perceptual,
title = {Perceptual adaptation to non-native speech},
journal = {Cognition},
volume = {106},
number = {2},
pages = {707-729},
year = {2008},
doi = {https://doi.org/10.1016/j.cognition.2007.04.005},
url = {https://www.sciencedirect.com/science/article/pii/S0010027707001126},
author = {Ann R. Bradlow and Tessa Bent},
}

@article{bradlow1997intelligibility,
title = {Intelligibility of normal speech I: Global and fine-grained acoustic-phonetic talker characteristics},
journal = {Speech Communication},
volume = {20},
number = {3},
pages = {255-272},
year = {1996},
note = {Acoustic Echo Control and Speech Enhancement Techniques},
doi = {https://doi.org/10.1016/S0167-6393(96)00063-5},
url = {https://www.sciencedirect.com/science/article/pii/S0167639396000635},
author = {Ann R. Bradlow and Gina M. Torretta and David B. Pisoni},
}

@article{cooke2008foreign,
  title={The foreign language cocktail party problem: Energetic and informational masking effects in non-native speech perception},
  author={Cooke, Martin and Garcia Lecumberri, ML and Barker, Jon},
  journal={The Journal of the Acoustical Society of America},
  volume={123},
  number={1},
  pages={414--427},
  year={2008},
  publisher={AIP Publishing}
}

@article{best1995direct,
  title={A direct realist view of cross-language speech perception},
  author={Best, Catherine T},
  journal={Speech perception and linguistic experience},
  volume={171},
  year={1995},
  publisher={York. Press}
}

@article{kreiman1993perceptual,
  title={Perceptual evaluation of voice quality: review, tutorial, and a framework for future research},
  author={Kreiman, Jody and Gerratt, Bruce R and Kempster, Gail B and Erman, Andrew and Berke, Gerald S},
  journal={Journal of Speech, Language, and Hearing Research},
  volume={36},
  number={1},
  pages={21--40},
  year={1993},
  publisher={American Speech-Language-Hearing Association}
}

@article{oates2009auditory,
  title={Auditory-perceptual evaluation of disordered voice quality},
  author={Oates, Jennifer},
  journal={Folia Phoniatrica et Logopaedica},
  volume={61},
  number={1},
  pages={49},
  year={2009},
  publisher={S. Karger AG}
}

@article{landis1977measurement,
  title={The measurement of observer agreement for categorical data},
  author={Landis, J Richard and Koch, Gary G},
  journal={biometrics},
  pages={159--174},
  year={1977},
  publisher={JSTOR}
}

@article{morales2020sensitivenets,
  title={SensitiveNets: Learning agnostic representations with application to face images},
  author={Morales, Aythami and Fierrez, Julian and Vera-Rodriguez, Ruben and Tolosana, Ruben},
  journal={IEEE Transactions on Pattern Analysis and Machine Intelligence},
  volume={43},
  number={6},
  pages={2158--2164},
  year={2020},
  publisher={IEEE}
}

@article{maryn2009acoustic,
  title={Acoustic measurement of overall voice quality: a meta-analysis},
  author={Maryn, Youri and Roy, Nelson and De Bodt, Marc and Van Cauwenberge, Paul and Corthals, Paul},
  journal={The Journal of the Acoustical Society of America},
  volume={126},
  number={5},
  pages={2619--2634},
  year={2009},
  publisher={AIP Publishing}
}

@book{maryn2010acoustic,
  title={Acoustic measurement of overall voice quality in sustained vowels and continuous speech},
  author={Maryn, Youri},
  year={2010},
  publisher={Ghent University}
}

@book{laver1981acoustic,
  title={The phonetic description of voice quality},
  author={Laver, John},
  year={1980},
  publisher={Cambdrige University Press}
}

@article{patel2018recommended,
  title={Recommended protocols for instrumental assessment of voice: American Speech-Language-Hearing Association expert panel to develop a protocol for instrumental assessment of vocal function},
  author={Patel, Rita R and Awan, Shaheen N and Barkmeier-Kraemer, Julie and Courey, Mark and Deliyski, Dimitar and Eadie, Tanya and Paul, Diane and {\v{S}}vec, Jan G and Hillman, Robert},
  journal={American journal of speech-language pathology},
  volume={27},
  number={3},
  pages={887--905},
  year={2018},
  publisher={American Speech-Language-Hearing Association}
}

@book{fant1970acoustic,
  title={Acoustic theory of speech production: with calculations based on X-ray studies of Russian articulations},
  author={Fant, Gunnar},
  number={2},
  year={1971},
  publisher={Walter de Gruyter}
}

@book{kummer2013speech,
  title={Cleft palate and craniofacial anomalies: the effects on speech and resonance},
  author={Kummer, Ann W},
  year={2001},
  publisher={Taylor \& Francis US}
}

@article{lunner2009cognition,
  title={Cognition and hearing aids},
  author={Lunner, Thomas and Rudner, Mary and R{\"o}nnberg, Jerker},
  journal={Scandinavian journal of psychology},
  volume={50},
  number={5},
  pages={395--403},
  year={2009},
  publisher={Wiley Online Library}
}

@article{rusz2021speech,
  title={Quantitative acoustic measurements for characterization of speech and voice disorders in early untreated Parkinson’s disease},
  author={Rusz, Jan and Cmejla, Roman and Ruzickova, Hana and Ruzicka, Evzen},
  journal={The journal of the Acoustical Society of America},
  volume={129},
  number={1},
  pages={350--367},
  year={2011},
  publisher={AIP Publishing}
}

@article{fagherazzi2021voice,
  title={Voice for health: the use of vocal biomarkers from research to clinical practice},
  author={Fagherazzi, Guy and Fischer, Aur{\'e}lie and Ismael, Muhannad and Despotovic, Vladimir},
  journal={Digital biomarkers},
  volume={5},
  number={1},
  pages={78--88},
  year={2021},
  publisher={S. Karger AG Basel, Switzerland}
}






\clearpage

\setcounter{table}{0}
\setcounter{figure}{0}
\setcounter{equation}{0}
\renewcommand{\tablename}{Supplementary Table}
\renewcommand{\figurename}{Supplementary Figure}
\renewcommand{\theequation}{S\arabic{equation}}

\section*{Supplementary information}

\section*{Supplementary Note 1: Speech anonymization methods and McAdams transformation}
\label{supnote:anonymization_method}

Speech anonymization aims to reduce the amount of speaker-identifying information contained in a speech signal while retaining the information needed for the intended downstream use. In clinical speech research, this downstream information may include linguistic content, perceptual abnormality, acoustic pathology markers, or machine-learning features for disorder classification. The present study does not propose a new anonymization algorithm. Instead, it evaluates the perceptual and clinical consequences of applying a fixed McAdams coefficient-based anonymization method to pathological speech, following our previous work on the same German clinical speech corpus \cite{arasteh2024addressing,arasteh2023effect,maier2009peaks}. This Supplementary Note summarizes the method used in the paper and places it in the broader landscape of speech anonymization approaches.

\subsection*{Families of speech anonymization methods}

Speech anonymization methods can broadly be grouped into deterministic signal-processing approaches and neural synthesis-based approaches \cite{NAUTSCH2019441,tomashenko2022voiceprivacy2022,tomashenko2022voiceprivacy,tomashenko2024voiceprivacy}. Signal-processing methods operate directly on the waveform or on classical acoustic representations derived from the waveform. They typically modify formant structure, pitch, spectral envelope, or timing through explicit mathematical transformations. Such methods are attractive when reproducibility, interpretability, and independence from training data are important. Neural synthesis-based methods, by contrast, usually decompose speech into linguistic, prosodic, and speaker-related representations, modify or replace the speaker representation, and resynthesize a waveform using a neural vocoder or text-to-speech model. These systems can provide strong privacy protection and high naturalness in some settings, but they often depend on large training corpora, learned speaker embeddings, pseudo-speaker construction, or target-speaker-like mappings.

The method used in the present study belongs to the signal-processing family. It is based on the McAdams coefficient transformation and modifies the spectral envelope estimated by linear predictive coding (LPC) \cite{mcadams1984spectral,patino2021speaker}. The method does not learn a model from the study data, does not require a speaker embedding, does not synthesize speech from text or discrete tokens, and does not map the input speaker to another target or pseudo-speaker. These properties were important for the present work, because the aim was to examine the perceptual consequences of a reproducible, non-neural, non-target-speaker anonymization transformation in a clinical speech cohort.

\subsection*{McAdams coefficient-based anonymization}

The McAdams anonymization method builds on the source-filter view of speech production, where the observed speech waveform is modeled as the output of a time-varying vocal-tract filter excited by a source signal. In this framework, the spectral envelope is associated with the vocal tract and contains formant information that contributes to perceived speaker identity. The residual excitation contains the part of the signal not explained by the LPC filter and preserves much of the original temporal and source-related structure.

For each short-time frame of an input waveform $x[n]$, LPC estimates an all-pole model of order $p$. The speech sample is approximated by a linear combination of previous samples plus a residual excitation term,

\begin{equation}
x[n] \approx \sum_{k=1}^{p} a_k x[n-k] + e[n],
\label{eq:sup_lpc_time}
\end{equation}
where $a_k$ denotes the LPC coefficient at lag $k$, $p$ is the LPC order, and $e[n]$ is the residual excitation. The corresponding prediction polynomial in the $z$-domain is

\begin{equation}
A(z) = 1 - \sum_{k=1}^{p} a_k z^{-k}.
\label{eq:sup_lpc_z}
\end{equation}
The roots of $A(z)$ define the poles of the all-pole vocal-tract filter. A complex pole can be written as

\begin{equation}
z_k = r_k e^{j\phi_k},
\label{eq:sup_pole}
\end{equation}
where $r_k$ is the pole magnitude and $\phi_k$ is the angular frequency. The McAdams transformation modifies the angular frequency of each complex pole according to

\begin{equation}
\phi_k' = \phi_k^{\alpha},
\label{eq:sup_mcadams_angle}
\end{equation}
where $\alpha$ is the McAdams coefficient. Values of $\alpha$ below or above 1 shift the formant structure in different directions by changing the angular spacing of the complex poles. The transformed pole is then reconstructed as

\begin{equation}
z_k' = r_k e^{j\phi_k'}.
\label{eq:sup_transformed_pole}
\end{equation}
Only poles with non-zero imaginary components are transformed; real poles are left unchanged. After transformation, the modified poles are converted back into LPC coefficients $\tilde{a}_k$. The anonymized waveform $\tilde{x}[n]$ is then reconstructed using the transformed LPC filter and the original residual excitation,

\begin{equation}
\tilde{x}[n] \approx \sum_{k=1}^{p} \tilde{a}_k \tilde{x}[n-k] + e[n].
\label{eq:sup_reconstruction}
\end{equation}
This procedure changes the spectral envelope and therefore shifts formant-related cues that contribute to perceived speaker identity. At the same time, it preserves the residual excitation, so temporal structure, rhythm, and much of the pitch trajectory are not explicitly replaced by a generated or target-speaker signal. The method is therefore fundamentally different from voice-conversion anonymization, where the original speaker representation is replaced by a different speaker representation and a new waveform is synthesized from that representation.

In the present study, the implementation followed the anonymization pipeline used in our earlier computational study on pathological speech \cite{arasteh2024addressing}. The LPC frame size, frame shift, LPC order, McAdams coefficient setting, and any randomization strategy for $\alpha$ were identical to that prior implementation. For every selected original utterance in the 180-speaker perceptual cohort, the algorithm generated one anonymized counterpart. These original-anonymized pairs were then used unchanged in the discrimination, quality-rating, and expert severity-rating experiments.

\subsection*{Neural synthesis-based anonymization methods}

Neural anonymization systems usually represent speech in a latent space in which linguistic content and speaker identity are at least partially separated. A common design first extracts content-related features, prosodic features, and a speaker representation from the input waveform. The speaker representation is then removed, replaced, averaged with other speaker embeddings, generated synthetically, or conditioned on an unrelated prompt. Finally, a neural synthesizer or vocoder generates a new waveform. These systems are often evaluated in the VoicePrivacy framework \cite{tomashenko2022voiceprivacy2022,tomashenko2022voiceprivacy,tomashenko2024voiceprivacy}, where automatic speaker-verification metrics quantify privacy and automatic speech-recognition or downstream-task metrics quantify utility.

One class of neural methods replaces the original speaker embedding with a pseudo-speaker embedding and then synthesizes the output waveform. In x-vector-based anonymization, the input waveform is processed to extract bottleneck linguistic features, pitch-related features, and an x-vector speaker embedding \cite{Snyderxvector,fang2019speaker,srivastava20_interspeech}. The original x-vector is replaced by a pseudo-speaker vector, often derived by selecting and averaging embeddings from an external pool of speakers that are distant from the original speaker under a speaker-similarity model. The anonymized waveform is then generated by a neural waveform model conditioned on the preserved linguistic and prosodic features and the substituted speaker representation. Conceptually, this can be written as

\begin{equation}
f_{\mathrm{BN}}, f_0, v_{\mathrm{spk}} = E(x[n]),
\label{eq:sup_xvector_extract}
\end{equation}

\begin{equation}
\tilde{v}_{\mathrm{spk}} = \frac{1}{N}\sum_{i=1}^{N} v_{\mathrm{pool}}^{(i)},
\label{eq:sup_xvector_average}
\end{equation}

\begin{equation}
\tilde{x}[n] = S(f_{\mathrm{BN}}, f_0, \tilde{v}_{\mathrm{spk}}),
\label{eq:sup_xvector_synthesis}
\end{equation}
where $E$ denotes feature extraction, $f_{\mathrm{BN}}$ denotes bottleneck linguistic features, $f_0$ denotes the pitch contour, $v_{\mathrm{spk}}$ denotes the original speaker embedding, $\tilde{v}_{\mathrm{spk}}$ denotes the anonymized pseudo-speaker embedding, and $S$ denotes the neural synthesis model. This family of methods can be powerful, but it changes identity by conditioning synthesis on a replacement speaker representation.

A related class uses generative models to create synthetic speaker embeddings before speech synthesis. In such systems, the input waveform is decomposed into phonetic content, duration, pitch, energy, and speaker identity features. The speaker embedding is replaced by a synthetic embedding, for example one generated by a generative adversarial network \cite{ganmain,Wassersteingan}, and a neural text-to-speech system \cite{10096607,meyer2023anonymizing} generates a mel-spectrogram that is converted to a waveform by a vocoder \cite{fastspeech,ren2020fastspeech2,kong2020hifigan,lux23_blizzard}. A schematic representation is

\begin{equation}
P, D, f_0, E_{\mathrm{energy}}, v_{\mathrm{spk}} = H(x[n]),
\label{eq:sup_tts_extract}
\end{equation}

\begin{equation}
M = F(P, D, \tilde{v}_{\mathrm{spk}}, \tilde{f}_0, \tilde{E}_{\mathrm{energy}}),
\label{eq:sup_tts_mel}
\end{equation}

\begin{equation}
\tilde{x}[n] = V(M),
\label{eq:sup_tts_waveform}
\end{equation}
where $P$ denotes phonetic content, $D$ denotes phone durations, $E_{\mathrm{energy}}$ denotes energy features, $\tilde{v}_{\mathrm{spk}}$ denotes a synthetic speaker embedding, $F$ denotes the acoustic model, $M$ denotes the generated mel-spectrogram, and $V$ denotes the neural vocoder. These systems anonymize speech by regenerating the waveform with a substituted identity representation, and therefore differ conceptually from the McAdams method used here.

A third family uses discrete token representations from self-supervised speech models or neural audio codecs \cite{Panariellospeak,Borsosnac,Chennac}. In these systems, the waveform is encoded into semantic tokens that represent linguistic content and acoustic tokens that capture waveform-level detail. Speaker-related acoustic style can be replaced or prompted with tokens from another source, after which a language model or decoder generates a new acoustic-token sequence and a codec decoder reconstructs the waveform \cite{hsu2021hubert,defossez2023encodec}. A simplified representation is

\begin{equation}
s = S_{\mathrm{sem}}(x[n]), \qquad a = S_{\mathrm{ac}}(x[n]),
\label{eq:sup_token_extract}
\end{equation}

\begin{equation}
\hat{a} = T(s, \tilde{a}_{\mathrm{prompt}}),
\label{eq:sup_token_generation}
\end{equation}

\begin{equation}
\tilde{x}[n] = D_{\mathrm{codec}}(\hat{a}),
\label{eq:sup_token_decode}
\end{equation}
where $s$ denotes semantic tokens, $a$ denotes acoustic tokens, $\tilde{a}_{\mathrm{prompt}}$ denotes acoustic prompt tokens from an alternative source, $T$ denotes a token-generation model, and $D_{\mathrm{codec}}$ denotes the codec decoder. This type of anonymization can strongly alter speaker identity, but it depends on learned tokenizers, generative models, and neural decoding.

Another synthesis-based approach uses vector-quantized bottleneck features to suppress speaker identity in the content representation before speaker-conditioned waveform synthesis \cite{champion2024anonymizingspeechevaluatingdesigning}. The input waveform is converted into vector-quantized linguistic features and prosodic features, while synthesis is conditioned on a selected speaker vector rather than the original speaker identity. In schematic form,

\begin{equation}
z_{\mathrm{VQ}}, f_0 = Q(x[n]),
\label{eq:sup_vq_extract}
\end{equation}

\begin{equation}
\tilde{x}[n] = S(z_{\mathrm{VQ}}, f_0, v_{\mathrm{target}}),
\label{eq:sup_vq_synthesis}
\end{equation}
where $z_{\mathrm{VQ}}$ denotes vector-quantized bottleneck features and $v_{\mathrm{target}}$ denotes the target or pseudo-speaker conditioning vector. This class also relies on synthesis conditioned on a replacement speaker representation, which makes it distinct from the deterministic waveform transformation evaluated in the present study.

\subsection*{Rationale for using McAdams anonymization in this study}

The central aim of this paper is not to benchmark multiple anonymization algorithms, but to determine whether a previously validated anonymization method preserves clinically relevant perceptual information in pathological speech. The McAdams method \cite{patino2021speaker,mcadams1984spectral} was selected because it is deterministic, reproducible, lightweight, and does not require the construction of pseudo-speakers or target-speaker mappings. These properties make it suitable for probing how a controlled formant-level transformation affects pathological speech perception across different clinical groups.
This design also creates an interpretable link between the anonymization mechanism and the expected perceptual consequences. Because the McAdams method mainly modifies the spectral envelope, it is expected to interact most strongly with perceptual features carried by formant and resonance structure. This is directly relevant to clinical speech, where different disorders emphasize different acoustic substrates. CLP and Dysglossia can involve resonance or articulatory changes, Dysarthria can involve articulatory, prosodic, and phonatory components, and Dysphonia primarily affects laryngeal voice quality. The same transformation may therefore have different perceptual consequences across groups, which is the main empirical question addressed in the paper.

The present work should therefore be read as a human-centered evaluation of one fixed anonymization transformation in clinical speech, not as a claim that McAdams anonymization is optimal for all clinical use cases. The method provides a controlled and interpretable test case for quantifying the trade-off between privacy protection, perceived signal quality, and preservation of pathological severity.

\clearpage

\begin{table*}[h]
\centering
\caption{Comprehensive summary of all statistical tests reported in the main Results. Tests are grouped by section and include the underlying unit of analysis, test statistic, exact $p$-value (uncorrected unless indicated), and any multiplicity correction applied. RM-ANOVA, repeated-measures one-way analysis of variance; MWU, two-sided Mann-Whitney $U$; FDR, Benjamini-Hochberg false discovery rate correction.}
\label{stab:all_tests}
\setlength{\tabcolsep}{5pt}
\renewcommand{\arraystretch}{1.08}
\scriptsize
\begin{tabular}{p{0.32\textwidth}p{0.2\textwidth}p{0.30\textwidth}p{0.08\textwidth}}
\toprule
Contrast & Unit of analysis & Test statistic & $p$-value \\
\midrule
\multicolumn{4}{l}{\textit{Discrimination, across groups}} \\
\midrule
Zero-shot accuracy across 6 groups   & Listener mean ($n=10$) & RM-ANOVA $F(5,45)=3.579$ & $p = 0.008$ \\
Few-shot accuracy across 6 groups    & Listener mean ($n=10$) & RM-ANOVA $F(5,45)=1.386$ & $p = 0.248$ \\
Per-listener learning effect (few $-$ zero) & Listener mean ($n=10$) & Paired two-sided $t$-test, $t=2.990$, $\Delta=+1.9$ pp & $p = 0.015$ \\
\midrule
\multicolumn{4}{l}{\textit{Discrimination, listener attributes}} \\
\midrule
Native vs non-native, zero-shot      & Cell-level ($n=30,30$) & MWU & $p = 0.014$ \\
Native vs non-native, few-shot       & Cell-level ($n=30,30$) & MWU & $p = 0.083$ \\
Expert vs non-expert, zero-shot      & Cell-level ($n=24,36$) & MWU & $p = 0.625$ \\
Expert vs non-expert, few-shot       & Cell-level ($n=24,36$) & MWU & $p = 0.643$ \\
\midrule
\multicolumn{4}{l}{\textit{Quality}} \\
\midrule
Original vs anonymized quality       & Listener mean ($n=10$) & Paired two-sided $t$-test, $t=16.515$ & $p < 0.001$ \\
Original quality across 6 groups     & Listener mean ($n=10$) & RM-ANOVA $F(5,45)=4.025$ & $p = 0.004$ \\
Anonymized quality across 6 groups   & Listener mean ($n=10$) & RM-ANOVA $F(5,45)=2.760$ & $p = 0.029$ \\
QDS across 6 groups                  & Listener mean ($n=10$) & RM-ANOVA $F(5,45)=7.690$ & $p < 0.001$ \\
QDS native vs non-native             & Cell-level ($n=30,30$) & MWU & $p = 0.108$ \\
QDS expert vs non-expert             & Cell-level ($n=24,36$) & MWU & $p = 0.008$ \\
\midrule
\multicolumn{4}{l}{\textit{Demographics}} \\
\midrule
Sex, zero-shot, full cohort          & Per-speaker ($n=93,87$) & MWU & $p = 0.716$ \\
Sex, few-shot, full cohort           & Per-speaker ($n=93,87$) & MWU & $p = 0.968$ \\
Sex, QDS, full cohort                & Per-speaker ($n=93,87$) & MWU & $p = 0.140$ \\
Age vs zero-shot, full cohort        & Per-speaker ($n=180$)  & Spearman $\rho = +0.072$ & $p = 0.339$ \\
Age vs few-shot, full cohort         & Per-speaker ($n=180$)  & Spearman $\rho = +0.138$ & $p = 0.065$ \\
Age vs QDS, full cohort              & Per-speaker ($n=180$)  & Spearman $\rho = -0.147$ & $p = 0.048$ \\
Age vs zero-shot, adults only        & Per-speaker ($n=120$)  & Spearman $\rho = +0.061$ & $p = 0.506$ \\
Age vs few-shot, adults only         & Per-speaker ($n=120$)  & Spearman $\rho = +0.064$ & $p = 0.487$ \\
Age vs QDS, adults only              & Per-speaker ($n=120$)  & Spearman $\rho = -0.302$ & $p = 0.001$ \\
\midrule
\multicolumn{4}{l}{\textit{Perceptual--computational alignment}} \\
\midrule
Zero-shot vs EER                     & Group mean ($n=4$) & Pearson $r = +0.028$ & $p = 0.972$ \\
Few-shot vs EER                      & Group mean ($n=4$) & Pearson $r = +0.092$ & $p = 0.908$ \\
QDS vs EER                           & Group mean ($n=4$) & Pearson $r = -0.546$ & $p = 0.454$ \\
Zero-shot vs $\Delta$AUROC           & Group mean ($n=4$) & Pearson $r = +0.719$ & $p = 0.281$ \\
Few-shot vs $\Delta$AUROC            & Group mean ($n=4$) & Pearson $r = +0.728$ & $p = 0.272$ \\
QDS vs $\Delta$AUROC                 & Group mean ($n=4$) & Pearson $r = +0.453$ & $p = 0.547$ \\
Zero-shot vs anonymized AUROC        & Group mean ($n=4$) & Pearson $r = -0.817$ & $p = 0.183$ \\
QDS vs anonymized AUROC              & Group mean ($n=4$) & Pearson $r = -0.897$ & $p = 0.103$ \\
\bottomrule
\end{tabular}
\end{table*}

\begin{table*}[h]
\centering
\caption{Cell-level discrimination accuracy and quality degradation score by listener and speaker group. The $10 \times 6$ matrices are the per-listener per-group means. Listener attributes (Lang: N native, NN non-native; Exp: E expert, NE non-expert) are repeated in the first two columns for convenience. CLP, cleft lip and palate; CA, control adults; CC, control children. Values are integer percentages; QDS is original quality minus anonymized quality on a 0--100 scale.}
\label{stab:cell_matrices}
\setlength{\tabcolsep}{6pt}
\renewcommand{\arraystretch}{1.08}
\begin{tabular}{@{}llcccccccc@{}}
\toprule
Listener & Lang/Exp & CLP & Dysarthria & Dysglossia & Dysphonia & CA & CC & Row mean \\
\midrule
\multicolumn{9}{l}{\textit{Zero-shot discrimination accuracy}} \\
\midrule
L1  & NN / NE & \phantom{0}80 & \phantom{0}90 & \phantom{0}80 & \phantom{0}77 & \phantom{0}73 & \phantom{0}87 & 81 \\
L2  & NN / E  & 100 & \phantom{0}97 & \phantom{0}90 & \phantom{0}83 & 100 & \phantom{0}97 & 94 \\
L3  & NN / E  & \phantom{0}80 & \phantom{0}90 & \phantom{0}80 & \phantom{0}70 & \phantom{0}73 & \phantom{0}83 & 79 \\
L4  & NN / NE & 100 & 100 & \phantom{0}90 & \phantom{0}93 & \phantom{0}97 & 100 & 97 \\
L5  & NN / NE & \phantom{0}63 & \phantom{0}90 & \phantom{0}90 & \phantom{0}93 & \phantom{0}77 & 100 & 86 \\
L6  & N  / NE & 100 & 100 & \phantom{0}93 & \phantom{0}83 & \phantom{0}97 & \phantom{0}93 & 94 \\
L7  & N  / NE & \phantom{0}77 & 100 & \phantom{0}90 & \phantom{0}83 & \phantom{0}90 & \phantom{0}93 & 89 \\
L8  & N  / E  & 100 & \phantom{0}97 & \phantom{0}93 & 100 & 100 & 100 & 98 \\
L9  & N  / NE & \phantom{0}97 & 100 & \phantom{0}93 & \phantom{0}87 & \phantom{0}97 & \phantom{0}97 & 95 \\
L10 & N  / E  & \phantom{0}87 & \phantom{0}93 & \phantom{0}87 & \phantom{0}93 & 100 & \phantom{0}97 & 93 \\
\midrule
Column mean & & \phantom{0}88 & \phantom{0}96 & \phantom{0}89 & \phantom{0}86 & \phantom{0}90 & \phantom{0}95 & \phantom{0}91 \\
\midrule
\multicolumn{9}{l}{\textit{Few-shot discrimination accuracy}} \\
\midrule
L1  & NN / NE & \phantom{0}60 & \phantom{0}93 & \phantom{0}90 & \phantom{0}87 & \phantom{0}87 & 100 & 86 \\
L2  & NN / E  & 100 & \phantom{0}97 & \phantom{0}93 & \phantom{0}87 & 100 & \phantom{0}97 & 96 \\
L3  & NN / E  & \phantom{0}87 & \phantom{0}93 & \phantom{0}87 & \phantom{0}77 & \phantom{0}70 & \phantom{0}77 & 82 \\
L4  & NN / NE & 100 & 100 & \phantom{0}93 & \phantom{0}93 & \phantom{0}97 & 100 & 97 \\
L5  & NN / NE & \phantom{0}80 & 100 & \phantom{0}93 & 100 & \phantom{0}73 & \phantom{0}93 & 90 \\
L6  & N  / NE & 100 & 100 & \phantom{0}93 & \phantom{0}93 & 100 & \phantom{0}97 & 97 \\
L7  & N  / NE & \phantom{0}90 & \phantom{0}93 & \phantom{0}87 & \phantom{0}80 & \phantom{0}93 & \phantom{0}83 & 88 \\
L8  & N  / E  & 100 & \phantom{0}97 & \phantom{0}93 & 100 & 100 & 100 & 98 \\
L9  & N  / NE & \phantom{0}97 & 100 & \phantom{0}93 & \phantom{0}90 & \phantom{0}97 & \phantom{0}97 & 96 \\
L10 & N  / E  & \phantom{0}97 & \phantom{0}97 & \phantom{0}93 & \phantom{0}93 & 100 & \phantom{0}97 & 96 \\
\midrule
Column mean & & \phantom{0}91 & \phantom{0}97 & \phantom{0}92 & \phantom{0}90 & \phantom{0}92 & \phantom{0}94 & \phantom{0}93 \\
\midrule
\multicolumn{9}{l}{\textit{QDS = original $-$ anonymized}} \\
\midrule
L1  & NN / NE & 38 & 32 & 21 & 18 & 46 & 35 & 32 \\
L2  & NN / E  & 38 & 31 & 25 & 15 & 40 & 32 & 30 \\
L3  & NN / E  & 18 & 26 & 19 & 12 & 24 & 21 & 20 \\
L4  & NN / NE & 33 & 40 & 24 & 31 & 42 & 36 & 34 \\
L5  & NN / NE & 21 & 28 & 24 & 26 & 21 & 40 & 27 \\
L6  & N  / NE & 41 & 28 & 26 & 22 & 35 & 25 & 29 \\
L7  & N  / NE & 38 & 51 & 35 & 34 & 40 & 28 & 38 \\
L8  & N  / E  & 28 & 22 & 25 & 22 & 28 & 18 & 24 \\
L9  & N  / NE & 38 & 47 & 34 & 25 & 44 & 40 & 38 \\
L10 & N  / E  & 37 & 33 & 29 & 24 & 41 & 29 & 32 \\
\midrule
Column mean & & 33 & 34 & 26 & 23 & 36 & 30 & 30 \\
\bottomrule
\end{tabular}
\end{table*}

\begin{table*}[h]
\centering
\caption{Complete pairwise contrasts across speaker groups with Benjamini-Hochberg FDR correction, including non-significant rows. The main text reports only contrasts with $q < 0.05$; this table lists all 15 pairs for each of the three perceptual outcomes for transparency. Raw $p$ values come from paired two-sided $t$-tests on the per-listener-per-group means ($n=10$ listeners). FDR is applied separately within each outcome (15 contrasts each).}
\label{stab:pairwise_full}
\setlength{\tabcolsep}{6pt}
\renewcommand{\arraystretch}{1.08}
\begin{tabular}{p{0.34\textwidth}cccc}
\toprule
Contrast & Zero-shot $p$ & Zero-shot $q$ & Few-shot $p$ & Few-shot $q$ \\
\midrule
CLP vs Dysarthria         & 0.055 & 0.165 & 0.140 & 0.470 \\
CLP vs Dysglossia         & 0.929 & 0.929 & 0.864 & 0.926 \\
CLP vs Dysphonia          & 0.662 & 0.709 & 0.819 & 0.926 \\
CLP vs Control-Adults     & 0.460 & 0.627 & 0.850 & 0.926 \\
CLP vs Control-Children   & 0.147 & 0.285 & 0.525 & 0.788 \\
Dysarthria vs Dysglossia  & $<$0.001 & 0.001 & $<$0.001 & $<$0.001 \\
Dysarthria vs Dysphonia   & 0.008 & 0.030 & 0.005 & 0.037 \\
Dysarthria vs Control-Adults     & 0.078 & 0.196 & 0.157 & 0.470 \\
Dysarthria vs Control-Children   & 0.576 & 0.665 & 0.193 & 0.484 \\
Dysglossia vs Dysphonia          & 0.298 & 0.447 & 0.363 & 0.681 \\
Dysglossia vs Control-Adults     & 0.544 & 0.665 & $>$0.999 & $>$0.999 \\
Dysglossia vs Control-Children   & $<$0.001 & 0.002 & 0.226 & 0.484 \\
Dysphonia vs Control-Adults      & 0.211 & 0.352 & 0.662 & 0.903 \\
Dysphonia vs Control-Children    & $<$0.001 & 0.001 & 0.051 & 0.255 \\
Control-Adults vs Control-Children & 0.152 & 0.285 & 0.428 & 0.714 \\
\midrule
\multicolumn{5}{l}{\textit{Quality degradation score (QDS)}} \\
\midrule
Contrast & \multicolumn{2}{c}{QDS $p$ (raw)} & \multicolumn{2}{c}{QDS $q$ (FDR)} \\
\midrule
CLP vs Dysarthria         & \multicolumn{2}{c}{0.751} & \multicolumn{2}{c}{0.751} \\
CLP vs Dysglossia         & \multicolumn{2}{c}{0.015} & \multicolumn{2}{c}{0.037} \\
CLP vs Dysphonia          & \multicolumn{2}{c}{0.007} & \multicolumn{2}{c}{0.020} \\
CLP vs Control-Adults     & \multicolumn{2}{c}{0.054} & \multicolumn{2}{c}{0.101} \\
CLP vs Control-Children   & \multicolumn{2}{c}{0.459} & \multicolumn{2}{c}{0.492} \\
Dysarthria vs Dysglossia         & \multicolumn{2}{c}{0.004} & \multicolumn{2}{c}{0.017} \\
Dysarthria vs Dysphonia          & \multicolumn{2}{c}{0.001} & \multicolumn{2}{c}{0.012} \\
Dysarthria vs Control-Adults     & \multicolumn{2}{c}{0.367} & \multicolumn{2}{c}{0.423} \\
Dysarthria vs Control-Children   & \multicolumn{2}{c}{0.252} & \multicolumn{2}{c}{0.315} \\
Dysglossia vs Dysphonia          & \multicolumn{2}{c}{0.062} & \multicolumn{2}{c}{0.103} \\
Dysglossia vs Control-Adults     & \multicolumn{2}{c}{0.004} & \multicolumn{2}{c}{0.017} \\
Dysglossia vs Control-Children   & \multicolumn{2}{c}{0.143} & \multicolumn{2}{c}{0.196} \\
Dysphonia vs Control-Adults      & \multicolumn{2}{c}{0.002} & \multicolumn{2}{c}{0.016} \\
Dysphonia vs Control-Children    & \multicolumn{2}{c}{0.021} & \multicolumn{2}{c}{0.045} \\
Control-Adults vs Control-Children & \multicolumn{2}{c}{0.084} & \multicolumn{2}{c}{0.126} \\
\bottomrule
\end{tabular}
\end{table*}

\begin{table*}[h]
\centering
\caption{Within-pathology speaker-sex contrasts on the three perceptual outcomes. Each row reports the per-speaker mean accuracy or QDS (averaged across the 10 listeners) for female and male speakers within one group, along with the two-sided Mann-Whitney $U$ $p$-value for the female-versus-male comparison. No contrast reaches $p < 0.05$ in any group, supporting the cohort-level conclusion that the McAdams transformation does not systematically advantage one sex over the other. The Dysphonia row is the most informative for this claim, given that the group's male-skewed composition (25 male, 5 female) creates the largest potential for masked or inflated apparent disparities. Values are mean $\pm$ SD, integer percentages.}
\label{stab:sex_contrasts}
\setlength{\tabcolsep}{6pt}
\renewcommand{\arraystretch}{1.08}
\begin{tabular}{lcccc}
\toprule
Group & Female / Male $n$ & Female (mean $\pm$ SD) & Male (mean $\pm$ SD) & MWU $p$ \\
\midrule
\multicolumn{5}{l}{\textit{Zero-shot discrimination accuracy}} \\
\midrule
CLP            & 19 / 11 & 87 $\pm$ 11 & 91 $\pm$ 10 & 0.323 \\
Dysarthria     & 13 / 17 & 98 $\pm$ \phantom{0}4 & 94 $\pm$ 12 & 0.448 \\
Dysglossia     & 16 / 14 & 89 $\pm$ 23 & 88 $\pm$ 24 & 0.759 \\
Dysphonia      & \phantom{0}5 / 25 & 80 $\pm$ 23 & 88 $\pm$ 14 & 0.621 \\
Adult controls & 20 / 10 & 90 $\pm$ 12 & 90 $\pm$ 11 & 0.778 \\
Child controls & 20 / 10 & 96 $\pm$ \phantom{0}6 & 93 $\pm$ 13 & 0.919 \\
\midrule
\multicolumn{5}{l}{\textit{Few-shot discrimination accuracy}} \\
\midrule
CLP            & 19 / 11 & 91 $\pm$ \phantom{0}9 & 91 $\pm$ \phantom{0}7 & 0.724 \\
Dysarthria     & 13 / 17 & 98 $\pm$ \phantom{0}4 & 96 $\pm$ \phantom{0}9 & 0.529 \\
Dysglossia     & 16 / 14 & 92 $\pm$ 20 & 91 $\pm$ 24 & 0.688 \\
Dysphonia      & \phantom{0}5 / 25 & 92 $\pm$ \phantom{0}8 & 90 $\pm$ 16 & 0.808 \\
Adult controls & 20 / 10 & 90 $\pm$ \phantom{0}8 & 94 $\pm$ \phantom{0}7 & 0.222 \\
Child controls & 20 / 10 & 96 $\pm$ \phantom{0}8 & 91 $\pm$ \phantom{0}9 & 0.097 \\
\midrule
\multicolumn{5}{l}{\textit{Quality degradation score (QDS)}} \\
\midrule
CLP            & 19 / 11 & 32 $\pm$ 13 & 34 $\pm$ 17 & 0.931 \\
Dysarthria     & 13 / 17 & 38 $\pm$ \phantom{0}8 & 31 $\pm$ 14 & 0.449 \\
Dysglossia     & 16 / 14 & 26 $\pm$ 21 & 27 $\pm$ 22 & 1.000 \\
Dysphonia      & \phantom{0}5 / 25 & 25 $\pm$ 16 & 22 $\pm$ 14 & 0.717 \\
Adult controls & 20 / 10 & 36 $\pm$ 13 & 37 $\pm$ 10 & 0.808 \\
Child controls & 20 / 10 & 30 $\pm$ 10 & 30 $\pm$ 13 & 0.791 \\
\bottomrule
\end{tabular}
\end{table*}

\end{document}